%% file: main.tex
\documentclass[acmsmall]{acmart}

\usepackage{booktabs}
\usepackage{vcell}
\usepackage{subfloat}
\usepackage{booktabs}
 \usepackage{arydshln}
 
\usepackage{amsmath}
\usepackage[normalem]{ulem}

\newcommand{\rnr}[1]{{\color{black}{#1}}}

\newcommand{\mrev}[1]{{\color{black}{#1}}}

\AtBeginDocument{%
  \providecommand\BibTeX{{%
    \normalfont B\kern-0.5em{\scshape i\kern-0.25em b}\kern-0.8em\TeX}}}

\setcopyright{acmcopyright}
\copyrightyear{2022}
\acmYear{2022}
\acmDOI{xx.xxxx/xxxxxxx.xxxxxxx}
\acmConference[CSCW '22]{This work will appear in CSCW '22: Proc. ACM Hum.-Comput. Interact}{November 03--05, 2022}{Woodstock, NY}
\acmBooktitle{This work will appear in CSCW '22: Proc. ACM Hum.-Comput. Interact,
  June 03--05, 2022, Woodstock, NY}
\acmPrice{15.00}
\acmISBN{978-1-4503-XXXX-X/18/06}

\begin{document}

\title[Measuring the Prevalence of Anti-Social Behavior]{Measuring the Prevalence of Anti-Social Behavior in Online Communities}

\author{Joon Sung Park}
\affiliation{%
 \institution{Stanford University}
 \streetaddress{353 Jane Stanford Way}
 \city{Stanford}
 \state{California}
 \country{USA}}
\email{joonspk@stanford.edu}

\author{Joseph Seering}
\affiliation{%
 \institution{Stanford University}
 \streetaddress{353 Jane Stanford Way}
 \city{Stanford}
 \state{California}
 \country{USA}}
\email{seeringj@stanford.edu}

\author{Michael S. Bernstein}
\affiliation{%
 \institution{Stanford University}
 \streetaddress{353 Jane Stanford Way}
 \city{Stanford}
 \state{California}
 \country{USA}}
\email{msb@cs.stanford.edu}

\renewcommand{\shortauthors}{Joon Sung Park et al.}

\begin{abstract}
    With increasing attention to online anti-social behaviors such as personal attacks and bigotry, it is critical to have an accurate accounting of how widespread anti-social behaviors are. In this paper, we empirically measure the prevalence of anti-social behavior in one of the world's most popular online community platforms. We operationalize this goal as measuring the proportion of unmoderated comments in the 97 most popular communities on Reddit that violate eight widely accepted platform norms. To achieve this goal, we contribute a human-AI pipeline for identifying these violations and a bootstrap sampling method to quantify measurement uncertainty. We find that 6.25\% (95\% Confidence Interval [5.36\%, 7.13\%]) of all comments in 2016, and 4.28\% (95\% CI [2.50\%, 6.26\%]) in 2020-2021, are violations of these norms. Most anti-social behaviors remain unmoderated: moderators only removed one in twenty violating comments in 2016, and one in ten violating comments in 2020. Personal attacks were the most prevalent category of norm violation; pornography and bigotry were the most likely to be moderated, while politically inflammatory comments and misogyny/vulgarity were the least likely to be moderated. This paper offers a method and set of empirical results for tracking these phenomena as both the social practices (e.g., moderation) and technical practices (e.g., design) evolve.

\end{abstract}



\maketitle


\input{content_minor_revision__Apr2022/introduction}

\input{content_minor_revision__Apr2022/related_work}
\input{content_minor_revision__Apr2022/method}
\input{content_minor_revision__Apr2022/method-pt2}
\input{content_minor_revision__Apr2022/method-pt3}

\input{content_minor_revision__Apr2022/results}

\input{content_minor_revision__Apr2022/discussion}
\input{content_minor_revision__Apr2022/conclusion}

\bibliographystyle{ACM-Reference-Format}
\bibliography{main}

\appendix
\input{content_minor_revision__Apr2022/appendix}

\end{document}

%% file: content_minor_revision__Apr2022/introduction.tex
\section{Introduction}
How widespread is anti-social behavior online? Surveys suggest that over four in ten U.S. adults have experienced harassment online~\cite{pew_harassment_results}, including the vast majority of women~\cite{vitak_harassment}, and that online political hostility is omnipresent~\cite{bor_hostility}. Over 3\% of posts on popular news websites are flagged by users for rule violations~\cite{cheng2017anyone}, and 96\% of minutes watched on a popular video streaming platform included at least one action by moderators to remove a piece of violating content~\cite{twitch_transparency_report}. Anti-social behaviors such as inflammatory comments \cite{1_Chandrasekharan}, hate speech~\cite{2_Donovan}, trolling~\cite{82_Claire, 83_Kayany}, and other ``undesirable'' comments~\cite{4_Chancellor, 5_Cheng, 6_Sood, 7_Chandrasekharan} not only hamper discussions in online communities~\cite{88_Kraut}, but also cause serious harm~\cite{84_Yavuz, 85_Wiener} and seed even more anti-social behavior~\cite{cheng2017anyone}. As platforms such as Reddit, Facebook, and Twitter have grown in size and consequence, concerns over the prevalence of anti-social behavior on these platforms have mounted~\cite{71_Laub}. Platforms responded by deploying thousands of paid moderators~\cite{8_Gillespie, 9_Roberts}, an array of algorithms~\cite{28_Gorwa, 11_Hosseini}, and tools for community self-moderation~\cite{seering2020reconsidering, chandrasekharan2019crossmod}. Researchers have likewise sought to help reduce the prevalence of these behaviors, for example by creating tools for governance~\cite{13_Fan,zhang2020policykit} and automatic detection of anti-social behaviors~\cite{14_Butler, 15_Xu}.

Given these harms, and the effort to combat them, it is critical to understand how much of the content in online communities remains anti-social. A large empirical foundation in social psychology demonstrates that if these behaviors are visible and widespread, it will encourage others to engage in anti-social behaviors as well~\cite{5_Cheng, cheng2017anyone}. Benchmarking progress, or regression, requires an honest accounting of the situation. Platforms themselves have begun measuring the prevalence of specific behaviors such as hate speech and harassment~\cite{58_Culliford, 77_YoutubeTeam, 78_Vincent}.

Despite this need, empirically measuring the prevalence of anti-social behavior remains difficult. AI tools remain too error-prone to be fully relied upon~\cite{10_Binns, 11_Hosseini}, and manual labeling via random sampling is labor-intensive to perform at scale. In addition, defining which behaviors cross the line remains contentious, with different online communities applying different definitions~\cite{71_Laub} and platforms each establishing different standards for content and enforcement~\cite{71_Laub, 79_Klonick}. Therefore, it is no surprise that empirical studies that measure the prevalence of anti-social behavior are few and far between, with platform-published performance metrics often tucked under the platforms’ “transparency” or compliance reports using vaguely defined categories~\cite{58_Culliford}. 

In this paper, we develop a method to measure the proportion of these behaviors in online communities, and apply that method to measure the prevalence of anti-social behaviors on a major platform. Our focus is on Reddit, one of the most popular websites in the United States, and one where interaction is spread across thousands of independent communities (subreddits). While norms vary across subreddit, we adapt a set of macro Reddit moderation norms, such as removing misogyny and hate speech, from prior work~\cite{Chandrasekharan2018internet} that identified a set of these macro-level norms for user behavior and moderation that are shared by a strong majority of the largest subreddits. We verify that these norms are enforced by moderators across essentially all of the most popular subreddits, and then set out to sample comments on these popular subreddits to identify how many unmoderated comments violate each of these norms. We also seek to measure how many of comments were removed by existing moderation tools and processes.

More precisely, we present a fully realized human-AI pipeline that identifies these macro-norm violating comments at scale, and a bootstrap estimation procedure for estimating overall prevalence rates based on the identified violating comments. We use a publicly available dataset of removed comments from the 100 most popular subreddits  \cite{Chandrasekharan2018internet}, and a matching dataset of comments that were not removed, to train classifiers that can identify macro norm violations with recall at $0.99$. To validate our classifiers' decisions and complement high recall with high precision, we employ crowd workers to annotate a sample of the comments that the classifiers flagged, confirming whether these comments indeed violate at least one of the eight macro norms previously identified. We then model this process via a statistical bootstrap to estimate confidence intervals for the proportion of violating comments that remain on popular subreddits. We repeat this analysis twice, once on a dataset from May 2016--March 2017 and once from the last three weeks of December 2020.

Using our approach, we find that that 6.39\% (95\% confidence interval [5.49\%, 7.39\%]) of all comments on the platform in 2016--2017 are macro norm violations, as are 4.44\% (95\% CI [2.61\%, 6.39\%]) of comments on the platform in December 2020. So, even relatively recently, roughly one in twenty comments is a violation of norms against behaviors such as misogyny, threatening violence, and hate speech. Overall, moderation only catches a small percentage of violating comments, removing 4.86\% of these violations in 2016--2017, and 10.54\% of the violations in 2020. Some categories of violation were more likely to slip through the cracks: politically inflammatory comments and misogyny/vulgarity were the least likely to be moderated, while pornography and bigotry (particularly in 2020) were among the most aggressively moderated. Subreddits about hobbies and occupations had the fewest violating comments remaining online, controlling for the number of comments submitted and the ratio of comments per moderator.

Our results highlight the prevalence of content that violates established macro norms: they imply that over 16 million macro norm violating comments from the May 2016--March 2017 period, and over a half a million macro norm violating comments from the last three weeks of December 2020, remain unmoderated. (Though not directly comparable, it raised an alarm in the content moderation community when Facebook's transparency report suggested that just 0.1\% of its content views are categorized as hate speech~\cite{58_Culliford}, which translates to millions of affected users). 
It is not surprising that nearly half of Americans and the vast majority of women experience harassment, when 4.44\% of comments that remain online are norm-violating: it suggests that a person would encounter anti-social behavior after reading 23 random comments on Reddit on average. Chances are high (96\%) that the comment would either be a personal attack or misogyny/vulgarity.

These violations are often less flagrant than highly publicized issues such as misinformation campaigns or hate speech, but they can aggregate into an unwelcoming atmosphere.
We offer a reflection of how our results might help platform designers and community members make decisions about their processes. 
Finally, we point to the further need for empirical measurements and auditing of our content moderation strategies, and provide a working model pipeline for identifying norm-violating behavior online. Our code is available open-source at \url{https://github.com/StanfordHCI/ContentModAudit_CodeRelease}, so other researchers can apply it to measure moderation outcomes in any Reddit community.

\medskip

\textsc{Content Warning: } {\it Please be advised that some of the example comments in this paper contain offensive language.}

%% file: content_minor_revision__Apr2022/related_work.tex
\section{Related Work}
In this section, we cover prior work in characterizing and identifing anti-social behavior in online communities. Despite the continued effort to identify harmful content online, the existing approaches face significant methodological challenges. This presents a fertile ground in which empirical results highlighting what today's content moderation fails to capture could provide value in guiding the future effort in content moderation.

\subsection{Anti-social behavior in online communities}
Anti-social behavior has been documented for essentially as long as online communities have been documented~\cite{Dibbell1993rape}. A review of the causes of, and responses to, such behavior, is the focus of~\citeauthor{kiesler2012regulating}~\cite{kiesler2012regulating}. One form of anti-social behavior is trolling, which is typically defined as intentional disruption of the community. Trolling is sometimes attributed to the online disinhibition effect~\cite{suler2004online}, where we engage in behavior online that we might not in face-to-face interaction. There exists a set of individuals who are dispositionally oriented toward troll behavior~\cite{buckels2014trolls}, and engage in anti-social behavior because they are redirecting internal feelings~\cite{varjas2010high} or do it for fun~\cite{shachaf2010beyond}. While such individuals are equally uncivil both online and online, they have more reach and visibility online~\cite{bor_petersen_2021}.

Another form of anti-social behavior is non-premediated, often referred to as flaming~\cite{83_Kayany,Kiesler1984social}. Evidence suggests that many online users can be tipped into engaging in flaming through a combination of mood and environmental signals~\cite{cheng2017anyone}. Human observers can predict with reasonable accuracy whether a discussion thread is going to end with hostility~\cite{zhang2018conversations}, but we are poor at estimating how others will react to our own comments~\cite{chang2020don}.

Facebook answered the increasing pressure related to hate speech on its platform with a brief transparency report suggesting, for the first time, that roughly 1 in 1,000 content views on its platform include what the platform considers to be hate speech \cite{58_Culliford}. Similarly, Reddit's Safety Team announced the impact of a certain form of moderation like banning of a toxic community on curbing toxic content \cite{78_Vincent} while YouTube provided a summary of the type of content the platform is trying to remove \cite{77_YoutubeTeam}.

Measuring the prevalence of these behaviors remains an open challenge. One thread of prior work has used survey methods to understand the prevalance of anti-social behaviors online (e.g.,~\cite{pew_harassment_results,vitak_harassment}). Another thread has used log data to observe when content is flagged or removed (e.g.,~\cite{cheng2017anyone,twitch_transparency_report}). The present research extends these efforts through directly sampling and coding online behaviors as violations. Such an approach allows us to avoid sampling and reporting biases, though it makes tradeoffs to do so. For example, not all violations are created equal, and survey methods will be better at capturing the impact of seeing a violation.

\subsection{Moderation}
Content moderation is a common response to anti-social behaviors. Content moderation strategies can be broadly categorized into human moderation and automated moderation. Where online forums and discussion boards were previously managed in large part by human moderators who would carefully sift through the content posted and remove those that go against a given community's standard \cite{27_Lampe}, online social media companies today are increasingly relying on automated tools to scale up the moderation effort \cite{28_Gorwa, 29_Roberts}. But despite the ongoing efforts to improve these strategies, both human moderation and automated moderation face challenges.

\subsubsection{Human Content Moderators}
Online content moderation strategies that rely on human moderators to control anti-social content involve dedicated moderators to manually inspect and act on such content. Although human moderators can often make more nuanced decisions than automated moderators by taking into consideration the context \cite{28_Gorwa, 30_MSB, 31_Lessig, alkhatib2019street} and ``behind-the-scenes'' norms of a local community that are less explicit \cite{Chandrasekharan2018internet}, they face significant challenges in ensuring that all content is reviewed in modern-day social media platforms where participants number in hundreds of millions to even billions \cite{32_Preece, 33_Williams}. Therefore, human moderators often take a reactive instead of a proactive approach in which the content that users share is first posted online and then reviewed once they have been flagged by others as anti-social \cite{9_Roberts, 34_Gillespie}. This results in anti-social content remaining visible to the users within the online community until it is reviewed by a moderator, or in the worst case, remain unaddressed indefinitely \cite{34_Gillespie}. 

 
\subsubsection{Automated Content Moderators}
To moderate the large volume of content that is shared on social media platforms, \rnr{there has been an increasing effort to create tools that can automatically identify and act on anti-social behavior \cite{39_Bickert, 40_Google, 41_Twitter, 92_Chandrasekharan}}. \rnr{Such tools range from simple word or source-ban lists that filter content based on block-listed words, URLs, and source IP addresses \cite{42_HN}, to more sophisticated AI tools that leverage machine learning and natural language processing techniques \cite{chandrasekharan2019crossmod, 93_Chancellor}}. \rnr{Particularly within the context of self-governed online communities, recent academic research has tried to both better understand what the online communities value and consider to be norm violating content that should be removed \cite{Chandrasekharan2018internet, 35_Seering}, and build tools and datasets that would help these communities implement automated ways for identifying and flagging the norm violating content~\cite{1_Chandrasekharan, chandrasekharan2019crossmod, 93_Chancellor}.}

And concurrent with the research that showed automated tools could be useful for efficiently capturing textual cyberbullying \cite{43_Dinakar, 44_Xu}, vulgarity and other anti-social behaviors online \cite{4_Chancellor, 1_Chandrasekharan, 5_Cheng}, large social media platforms started to transition their moderation strategy to be more reliant on automated tools \cite{35_Seering, 45_Geiger}. For example, following a major public controversy in which Facebook reportedly became the medium for hate speech that incited genocide against the Muslim population in Myanmar, the social media platform improved its Burmese hate speech classifier to increase automated takedowns of content by 39\% \cite{46_GIFCT}. Meanwhile, YouTube reports that 93\% of the videos on its platform that are removed for violent extremism are flagged by machine-learning algorithms \cite{40_Google}, while Reddit’s subcommunities now routinely employ AutoModerator to automatically filter out comments violating the community standards \cite{47_reddit}.

However, the automated tools for fighting anti-social content online have received criticism for their limitations. One of the most commonly cited limitations includes automated moderation tools’ inability to take into consideration the context in which an online conversation is taking place \cite{30_MSB, 31_Lessig, 48_Pater}; \rnr{a comment may be construed as acceptable or even encouraged in one community while being completely unacceptable in another \cite{seering2020reconsidering}}. At the same time, many of the modern automated moderation tools are considered to be too brittle, often suffering from high false-positive rates and failing to understand the finer nuances or irony in user-generated content \cite{49_Sood, 50_Brody, 51_Chancellor, 52_li}. As a case in point, Jigsaw---a subsidiary of Google---published a Perspective API that aimed to provide a state of the art machine learning model with high test accuracy for quantifying ``toxicity'' of textual content. The API often failed to match observers’ expectations and could easily be tricked (e.g., the single-term comment ‘Arabs’ was determined to be 63\% toxic, while ‘I love fuhrer’ was only 3\% toxic \cite{53_Sinders}). Additionally, some worry that these automated tools would exhibit discriminatory bias like other machine learning models, thereby making moderation decisions that unfairly treat people of a certain gender, race, ethnicity, or age and cause more harm than good \cite{54_Diaz, 55_Blodgett, 56_Zehlike, 57_Barocas}. As such, there is a growing consensus that automated moderation tools should be used in conjunction with human moderators, for example by triaging potentially anti-social behavior for human moderators to review \cite{28_Gorwa, chandrasekharan2019crossmod}. 

Given this, rigorous empirical study exploring the prevalence of antisocial behavior, and the impact of content moderation, can help moderators and designers plan their strategies more effectively. Therefore, in this work, we aim to operationally define, quantify, and characterize anti-social behavior that populates some of the most popular communities on Reddit. 

%% file: content_minor_revision__Apr2022/method.tex
\section{Identifying Macro Norm violating comments}
Our aim is to identify macro norm violating comments on Reddit, to quantify their prevalence, and to characterize their content, rate of engagement, and language usage. However, there are too many comments for manual annotation, and state-of-the-art machine learning classifiers are not robust enough on their own. We overcome these issues using a human-AI pipeline in which we use classifiers with a high recall to nominate candidate comments that might be violating, and then focusing our manual annotations with trained annotators on these nominated comments. By tuning the classifiers to have high recall over high precision, our pipeline ensures that almost all of the violating comments are sent to be reviewed by our annotators. Additionally, in concurrence with prior work \cite{chandrasekharan2019crossmod, seering2020reconsidering}, this pipeline ensures that a human has the final say in labeling any piece of content as a violation.

In this section, we first discuss the scope of our investigation, including how we define violating comments for this paper. We then summarize our pipeline for identifying violating comments. 

\subsection{Scope of the Study}

\input{content_minor_revision__Apr2022/table/macro_norms}

Reddit, the focus of our study, is a large-scale social media platform with 52 million daily active users \cite{59_Phan}. On Reddit, users join smaller subcommunities called subreddits that cover a specific topic and are managed by voluntary moderators who enforce community-specific rules (e.g., the type of allowed content, expected member behaviors), making the platform a good test-bed for studying user behaviors across diverse sets of moderation strategies and topics. In particular, we explore comments posted in response to top-level post submissions as most of the discussions on Reddit take place in the comment section. Given that the subreddits each have varying rules, we consider a comment to be violating if it breaks one of the \textit{macro norms} on Reddit --- norms that the vast majority of subreddits agree on. These norms were identified in prior work~\cite{Chandrasekharan2018internet} that investigated the 100 most popular subreddits harboring nearly a third of all comments on Reddit to extract the topic categories for the moderated (summarized in Table 1). 

We rely on macro norms as they provide us with a lens to measure the prevalence of violating comments that are largely independent of community-relevant contexts. However, in doing so, we are explicitly not accounting for comments that do not violate macro norms but still violate local rules of the respective subreddit. We also note that some subreddits have explicitly chosen to permit content that violates one or more of these macro norms (e.g., vulgar or sexualized comments). This highlights an important tension between the local and the macro norms. We discuss these issues and expand on their implications for future content moderation in the discussion section.

\subsection{Data for Training and Testing the Classifiers}
The first step of our pipeline uses a set of machine learning classifiers to flag comments that are potentially macro norm violations. We trained and tested these classifiers following best practice by constructing a balanced dataset that contains an equal number of moderated comments (denoted as $\mathcal{M}$) and unmoderated comments that are still online and were not moderated or deleted by the author (denoted as $\mathcal{M'}$). There are more unmoderated than moderated comments on Reddit---$\mathcal{M'}$, therefore, is a subset of all unmoderated comments. While such balanced datasets do not match the real-world distribution, balancing the dataset gives the resulting model equal priority to each class, which is important for ensuring that our model actually learns the meaningful features for the classification task and not just the uneven class distribution. 

\subsubsection{\rnr{$\mathcal{M}$: moderated comments}} 
$\mathcal{M}$ represents the top 100 most popular English subreddits during the 11 month period from May 2016 to March 2017. Given that the moderated comments are removed soon after they are posted, prior work used \textit{praw}, a Reddit streaming API, to stream and save all comments posted to each of these study subreddits before they were moderated \cite{Chandrasekharan2018internet}. 24 hours after each of these comments were streamed, all comments were queried again via the API using their unique \textit{comment\_ID} and verified which of them were replaced by a [``removed''] tag as that would signal their removal due to moderation. Any comments by AutoModerator accounts, which are bots for moderation, were removed from $\mathcal{M}$. This left $\mathcal{M}$ with a total of 2,831,664 removed comments, with at least 5,000 for each of the 100 study subreddits.

\subsubsection{\rnr{$\mathcal{M'}$: unmoderated comments}} 
As $\mathcal{M}$ contained only the moderated comments from the sampling period, we collected $\mathcal{M'}$ ourselves through historical archives of Reddit comments. Of the 100 study subreddits from $\mathcal{M}$, three--- r/The\_Donald, r/Incels, r/soccerstreams---no longer exist on the platform, so we focused our investigation on the remaining 97 study subreddits. During the construction of $\mathcal{M'}$, we aimed to closely replicate the data collection process of $\mathcal{M}$. For each of the 97 subreddits, we used \textit{Pushshift} dataset that stores all content posted on the Reddit platform to gather IDs of submissions that were posted from the same timeframe as when $\mathcal{M}$ was collected with an even distribution across the 11 months. We then used \textit{praw} to get the actual comments with the submission IDs and discarded any that were posted by a bot or moderated. We continued this process until we had a balanced dataset for each of the 97 study subreddits.

\input{content_minor_revision__Apr2022/figure/pipeline_figure}

\subsection{Building the Classifiers}
Using $\mathcal{M}$ and $\mathcal{M'}$, we built 97 neural network binary classifiers, each of which was trained on the data from one of the 97 study subreddits to classify whether a given comment would be moderated on that subreddit. These classifiers collectively determine whether a comment is likely to have violated one of the macro norms and thus would have been removed on most of the subreddits. We refer to these classifiers as subreddit classifiers. 

\subsubsection{Preprocessing the data} 
We first preprocessed our dataset by putting all characters in lowercase and removing non-alphabetical characters. We then segmented our dataset into a \textit{training} dataset (70\% of all data) and \textit{testing} and \textit{validation} datasets (15\% of all data each), each with an equal number of moderated and unmoderated comments. For every study subreddits, we then used our training dataset to train word embeddings from scratch and encoded comments as fixed-length vectors, trancating and padding as needed. 

\subsubsection{Building the classifiers} 
We built our classifiers using Google's \textit{TensorFlow} and trained and validated them with the encoded dataset. The classifiers have a four-layer neural network architecture, starting with an embedding layer that takes the encoded list of integers and finds an embedding vector for each word, which we learned as we trained our network. We then pass through an average pooling layer that returns a fixed-length output vector and then through a dense layer with Rectified Linear Unit (ReLU) activation function \cite{26_TensorFlow}. Finally, we employ another dense layer with a sigmoid activation function that transforms the final output of the network into a value between 0.0 and 1.0. For our binary classification task, we identify a comment as one that would have been removed in a given subreddit if the final output of the network is greater than or equal to 0.5. 

We then fine-tuned the following four parameters for each of our subreddit classifiers using grid search where we try out exhaustive combinations of hyperparameters given candidate values:  the size of the word index used in the encoding, the length of the input vector, the number of epochs during the training phase, and the number of nodes for the ReLU layer of the neural network. These are summarized in Table 2. We optimized for the F1 score (\textit{f}-measure) on our validation dataset, achieving an average of 72.3 (std=4.32) across the 97 classifiers. This is comparable to the classifiers presented in prior work that were trained on a similar dataset \cite{4_Chancellor, chandrasekharan2019crossmod}.

\input{content_minor_revision__Apr2022/table/cf_parameters}

\subsection{Machine Learning Flags Comments}
We marked a comment as \textit{flagged} (potentially norm violating) if the number of subreddit classifiers that flagged the comment, which we call the \textit{classifier agreement score}, was greater than or equal to 80 out of 97. This threshold was selected to achieve a high recall on the ensemble classifier even at the cost of producing false positives as our pipeline includes human annotators who validate the classifier flagged comments. In other words, we wanted our process to miss as few violating comments as possible, so we deliberately used a low threshold of 80 out of 97 and passed these comments to a human review stage. This approach provides statistical power even within a realistic budget for manually annotating comments, because it results in roughly one in five flagged comments later being coded as a violation while maintaining a near-zero false negative rate.

We confirmed that this threshold indeed captures most of the violating comments: the first author manually annotated a random sample of 1,000 comments in the validation dataset from 2016 to 2017 period. This sample contained 400 comments with $< 80$ subreddit classifier agreement, 200 with $80 \leq \text{agreement} < 85 $, 200 with $85 \leq \text{agreement} < 90$, and 200 with $\geq 90$ agreement. We find that only one percent of the comments with $< 80$ classifier agreement violated at least one of the macro norms when manually inspected, whereas this number significantly increased in the subsequent sample groups (9.5\%, 12\%, and 28\% in the order of increasing classifier agreement). This low false negative rate when using a low enough agreement threshold matches the observations in a prior work that took a similar approach to classifying violating comments on Reddit \cite{chandrasekharan2019crossmod}.

In addition, to confirm that our classifier's low false negative rate holds for the comments from 2020 period, we further annotate 400 comments with classifier agreement score of less than 80 from this period randomly sampled across the study subreddits. We find our result to replicate, with roughly the same rate of 1.25\% of the sample to violate one of the macro norms. Although our subreddit classifiers were trained on comments from a 2016 to 2017 period, this low false negative rate for the comments from 2020 suggests that when combined with our human annotators, our overall pipeline still remains robust even for the newer comments. Finally, as we describe in the following section on our bootstrap sampling methods, these false negative rates are accounted for in our calculation of the confidence interval of our estimations.

\subsection{Human Annotation Validates the Flagged Comments}
\label{sec:annotation}
The tradeoff for tuning our classifiers for very few false negatives is that they produce more false positives. So we recruited human annotators to verify that the classifier flagged comments are indeed violating by asking them to code a subset of the flagged comments to see which macro norms they violate, where the subset was a random sample of the flagged comments with an even distribution across the 97 study subreddits. The definitions of these macro norms that we presented to our annotators were inspired by prior work~\cite{Chandrasekharan2018internet}, but based on our qualitative annotation discussed above, we found it appropriate to expand the definitions for some of them to better fit our data. We updated the norm described as ``opposing political views around Donald Trump'' to ``inflammatory political claims'' that covers inflammatory comments that are against the right-leaning and the left-leaning political ideologies and updated the norm described as ``hate speech that is racist or homophobic'' to ``bigotry'' that covers hate speech directed at ethnic or religious groups as well.

\input{content_minor_revision__Apr2022/figure/human_annotation_screenshots}

\subsubsection{Recruiting  crowd workers} 
The crowd workers were recruited from Amazon Mechanical Turk (MTurk), and they had to be at least 18 years old, living in the US, and have completed more than 1,000 Human Intelligence Tasks (HITs – MTurk’s task unit) with the minimum HIT approval rating of 98\%. In our pilot annotation task that we will describe in a subsequent subsection, our annotators took around 10 seconds (median=9.66 seconds; 75th percentile=16.99 seconds) to annotate a single comment. Based on this, we decided to pay our workers \$1.50 for every 30 comments they annotated to ensure that we are paying the majority of our workers at the rate of at least \$15.00 per hour. We decided on this rate informed by Rolf's \textit{The Fight For Fifteen} \cite{20_Rolf}.

\subsubsection{Human annotation workflow} 
Applying human annotation for non-trivial classification tasks could suffer from inaccurate annotations due accidental errors~\cite{22_Angeli, 23_Pershina, 24_Zhang}. Therefore, we designed a training and a testing phase that are inspired by the \textit{gated instruction} workflow~\cite{25_Liu} as follows to ensure that our annotators clearly understand and are proficient at the task:

Our annotators were directed to a custom-built web platform to which they could sign in with their MTurk ID. For those joining for the first time, they were redirected to the first portion of the training phase in which they were presented with 1) an overview description of the task and its goal, 2) a content warning notifying them that some comments in the task might include offensive language, and 3) the eight macro norms accompanied by their definitions and two gold-standard example comments that we manually chose from the test dataset (Figure 2-A). When they finished reviewing this content, they were asked to practice annotating 30 hand-selected, gold-standard examples of the macro norm violations handpicked from our test dataset. This was done on the actual interface used for the main annotation task that showed a comment to be annotated, and multi-select HTML form for submitting macro norm violations (Figure 2-B). The annotators could select any number of macro norms they thought the given comment violated. Importantly, during this training phase, the workers were presented with the correct annotation and explanation after each time they submitted their annotations for a given comment. Note also that, in order to avoid biasing their decisions by imposing an ``expert'' AI opinion, the workers were not told that the comments were flagged by an algorithm as potentially norm violating \cite{seering2020reconsidering}.

Of the 30 practice comments, the last 10 were effectively the test annotations; we measured our annotators’ Cronbach's alpha reliability score when compared to our gold-standard annotations, and only admitted those whose reliability score was greater than or equal to 0.7 for the last 10 training annotations. All annotators who participated in the training and testing phase were provided with a completion code they could submit to MTurk and were paid \$1.50 for their time. Those who were admitted were allowed to start the main annotation task. They could annotate as many comments as they wanted as long as there were more comments to be annotated, and were provided with a completion code they could submit to receive \$1.50 for every 30 comments they annotated. Each of the comments was seen by three unique annotators to test for majority agreement. 

During the course of this study, we recruited a total of 31 annotators who collectively annotated a total of 4,850 comments. \mrev{Finally, we followed best practices for accounting for annotator well-being during their task. In addition to presenting the aforementioned content warning, we made sure that our annotators could freely leave the study at any time if they felt uncomfortable with the annotation task. We purposefully designed our compensation scheme to ensure that we paid our participants in small increments instead of asking them for a long period of participation before receiving their compensation to ensure that our participants could receive the payment they deserve regardless of when they choose to leave the study.}

\subsubsection{Pilot} 
To confirm the robustness of this approach and  to determine the right level of compensation for the workers, we ran a pilot annotation task for which we recruited 20 annotators to partake in the training phase. Of the twenty, 8 annotators passed the testing phase and went on to annotate 194 randomly selected comments from the test dataset. The first author manually annotated the 194 comments that the annotators annotated during the aforementioned pilot tasks to establish baseline annotations to which annotators' work would be compared. In relation to the first author’s manual annotation, the majority-agreed annotation of the annotators yielded a Cronbach's alpha score of 0.86. 

%% file: content_minor_revision__Apr2022/table/macro_norms.tex
\begin{table}[tb]
\centering
\caption{The eight macro norm violations on 97 popular subreddits and their definitions. We took the norms uncovered in a prior study \cite{Chandrasekharan2018internet} and expanded on some of their definitions to better fit our data.}
\begin{tabular}{ll}
\multicolumn{1}{c}{\textbf{MACRO NORM VIOLATIONS}}                                                & \multicolumn{1}{c}{\textbf{EXAMPLE COMMENTS}}                                                                                                                                         \\ 
\toprule
\vcell{Using misogynistic or vulgar slurs}                                                        & \vcell{\textit{"god... I want sage to knock this c*** out"}}                                                                                                                          \\[-\rowheight]
\printcelltop                                                                                     & \printcelltop                                                                                                                                                                         \\
\vcell{Inflammatory political claims}                                                             & \vcell{\begin{tabular}[b]{@{}l@{}}\textit{"a day old troll complaining about liberals -- I }\\\textit{smell a lost trumpkin"}\end{tabular}}                                           \\[-\rowheight]
\printcelltop                                                                                     & \printcelltop                                                                                                                                                                         \\
\vcell{Bigotry}                                                                                   & \vcell{\begin{tabular}[b]{@{}l@{}}\textit{"punishment for not being hateful enough and }\\\textit{not destroying the gays"}\end{tabular}}                                             \\[-\rowheight]
\printcelltop                                                                                     & \printcelltop                                                                                                                                                                         \\
\vcell{\begin{tabular}[b]{@{}l@{}}Verbal attacks on Reddit or specific \\subreddits\end{tabular}} & \vcell{\begin{tabular}[b]{@{}l@{}}\textit{"also reddit sucks because a user making~an~error }\\\textit{refuses to delete their post and redo it }\\\textit{correctly"}\end{tabular}}  \\[-\rowheight]
\printcelltop                                                                                     & \printcelltop                                                                                                                                                                         \\
\vcell{Posting pornographic links}                                                                & \vcell{\textit{[URL]}}                                                                                                                                                                \\[-\rowheight]
\printcelltop                                                                                     & \printcelltop                                                                                                                                                                         \\
\vcell{Personal attacks}                                                                          & \vcell{\textit{"you know man youre kind of a f***ing d*****"}}                                                                                                                        \\[-\rowheight]
\printcelltop                                                                                     & \printcelltop                                                                                                                                                                         \\
\vcell{Abusing and criticizing moderators}                                                        & \vcell{\textit{"the mods in this sub need to wake the f*** up"}}                                                                                                                      \\[-\rowheight]
\printcelltop                                                                                     & \printcelltop                                                                                                                                                                         \\
\vcell{\begin{tabular}[b]{@{}l@{}}Claiming the other person is too \\sensitive\end{tabular}}      & \vcell{\textit{"get off the internet with your sensitive ass"}}                                                                                                                       \\[-\rowheight]
\printcelltop                                                                                     & \printcelltop                                                                                                                                                                         \\
\bottomrule
\end{tabular}
\end{table}

%% file: content_minor_revision__Apr2022/figure/pipeline_figure.tex
\begin{figure}[tb]
  \centering
  \includegraphics[width=0.90\textwidth]{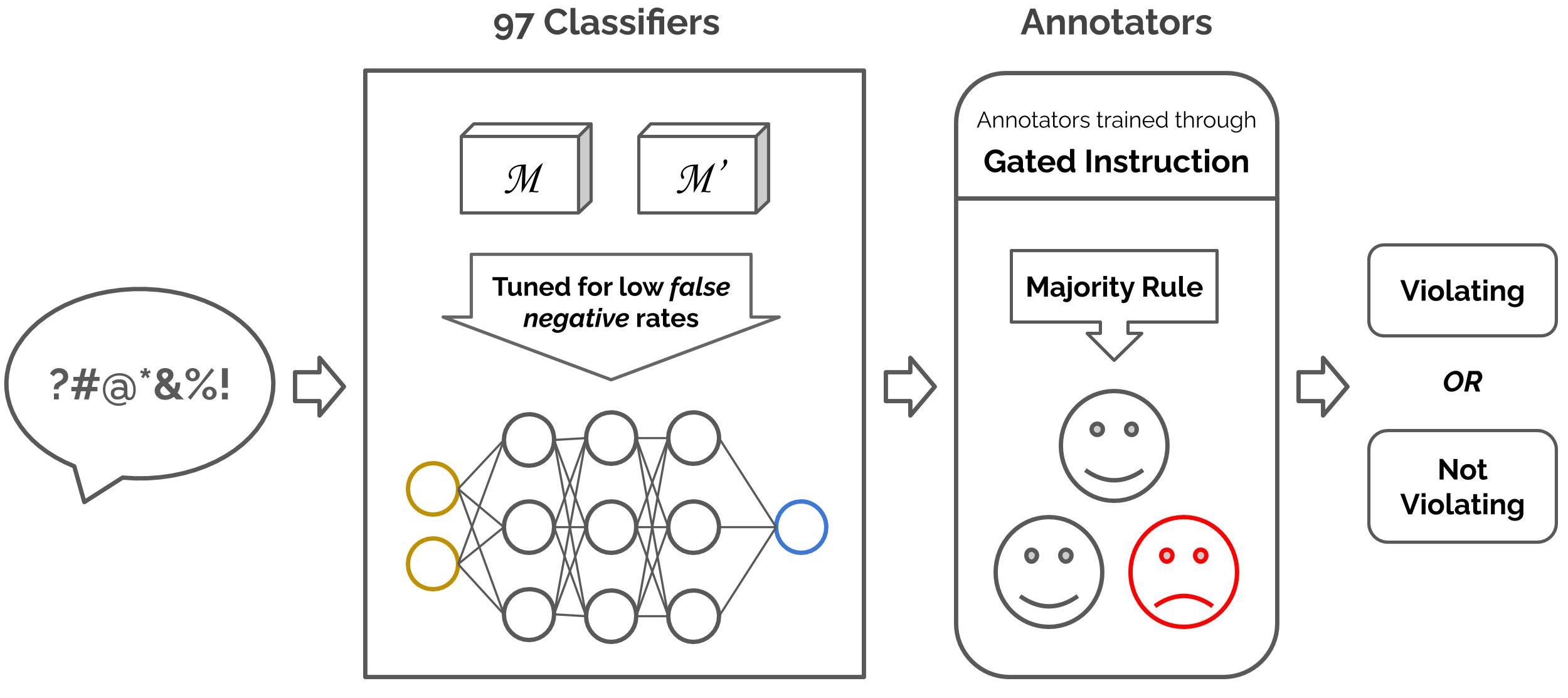}
  \caption{An illustration of the human-AI pipeline for identifying violating comments. Our pipeline includes 97 subreddit classifiers that are trained using a balanced dataset of moderated and unmoderated comments, and human annotators who are trained through gated instruction~\cite{25_Liu}. Our classifiers (tuned for high recall) nominate potentially violating comments and human annotators make the final determination.}
  \Description{Human-AI pipeline}
\end{figure}

%% file: content_minor_revision__Apr2022/table/cf_parameters.tex
\begin{table}[tb]
\centering
\caption{Parameters and the values used for them to fine-tune the classifiers}
\begin{tabular}{ll}
\multicolumn{1}{c}{\textbf{DESCRIPTION} } & \multicolumn{1}{c}{\textbf{SET OF VALUES} }  \\ 
\toprule
Size of the word index                    & {[}10000, 44000]                             \\
Max length of the input                   & {[}256, 512]                                 \\
Number of epochs during the training      & {[}30, 40, 50]                               \\
\# of nodes for the ReLU layer            & {[}16, 32]                                   \\
\bottomrule
\end{tabular}
\end{table}


%% file: content_minor_revision__Apr2022/figure/human_annotation_screenshots.tex
\begin{figure}[tb]
  \centering
  \includegraphics[width=0.98\textwidth]{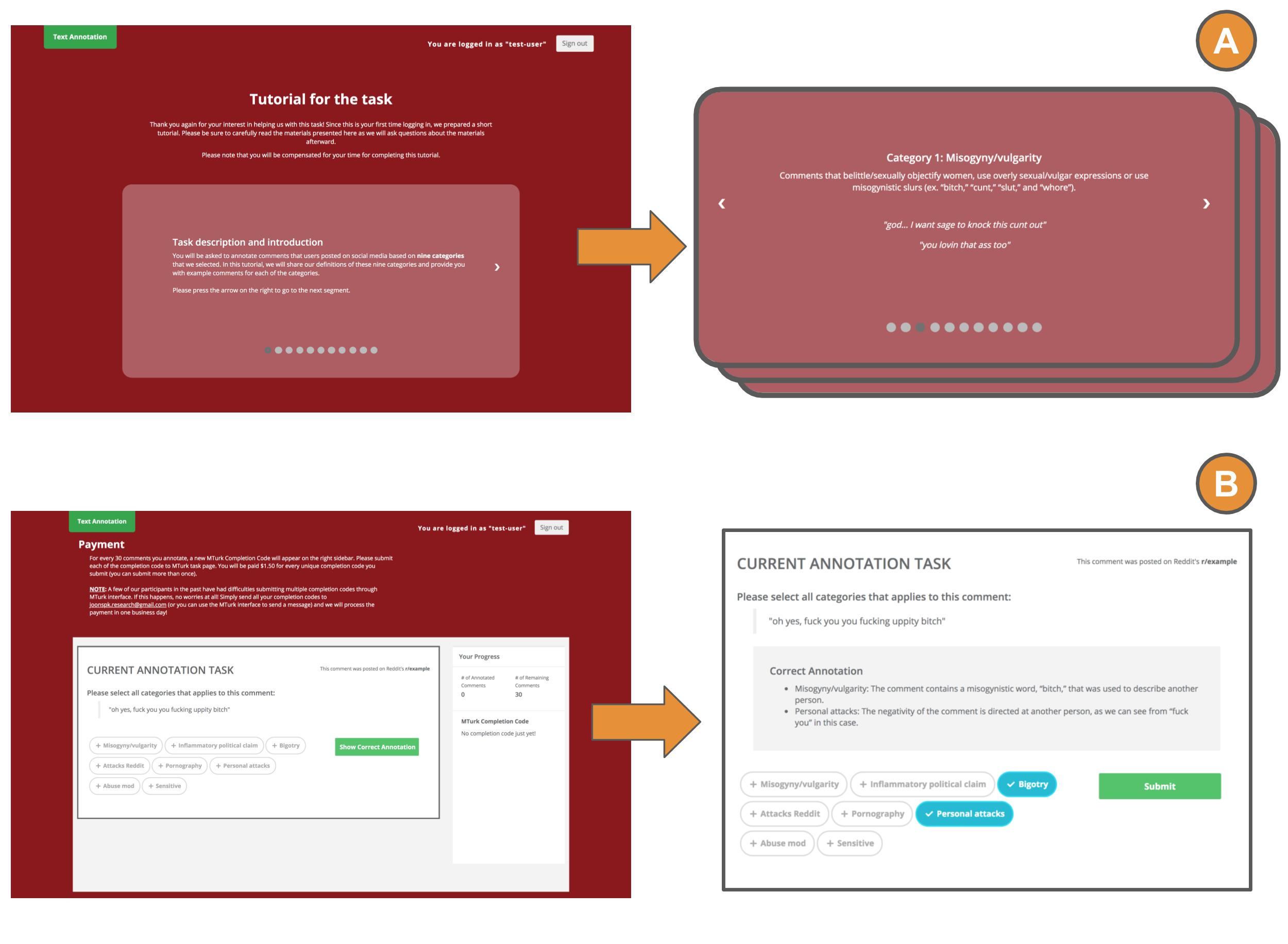}
  \caption{\textbf{A:} The interface for introducing the task description and eight macro norms to a new annotator. The definitions for each norms are shown one by one, accompanied by gold-standard examples that violate the norm. \textbf{B:} The interface for training and testing new annotators. Once the new annotators select their annotation, the correct annotation is shown accompanied by gold-standard examples. The interface for the main annotation task is the same but without presenting the correct annotation portion.}
  \Description{Annotation Interface}
\end{figure}

%% file: content_minor_revision__Apr2022/method-pt2.tex
\section{Bootstrap Sampling Reddit Comments for Analysis}
Having established the process for identifying macro norm violating comments on Reddit, we proceeded to apply this process to study the prevalence and characteristics of such comments. Our core strategy for doing this was to take random samples of the online comments from our study subreddits, calculating the classifier agreement scores for each of the samples’ elements, and then taking random samples from those comments that were flagged as potentially norm violating (classifier agreement $\geq 80$). But this effectively meant that we were sampling from different subsets of the population in which some of our random samples have a lower rate of violating comments than others due to the differences in the rate of violating comments between different subreddits. 

This makes drawing conclusions from our samples using the traditional inferential statistics problematic---we cannot simply calculate a binomial proportion confidence interval, because we have several convolved sources of uncertainty. The most straightforward parametric statistical procedure would be to select random comments, label them as violating or not violating, and estimate overall levels of violation from that. Unfortunately, due to the large class imbalance (most comments are not violating), this procedure is not tractable. Our introduction of the machine learning layer to nominate possible violations helps manage this problem, but threatens the random sampling procedure and can make errors itself. So, ultimately, we chose to use the classifiers to identify (noisily) a proportion of comments that are violating, complemented with human labeling at a smaller scale to verify. This means that our sampling procedure compounds multiple types of uncertainty: which comments are sampled from the dataset, which comments are flagged by the classifiers, and which comments are verified as actually violating by human annotators.

Given this, we applied a statistical \textit{bootstrapping} technique, variations of which have been used in prior work with similar compounded uncertainty, to derive an accurate measurement and confidence intervals. The core purpose of bootstrapping is to draw conclusions about a population by resampling with replacement from the sample data, which allows for direct observation of the sampling distribution of statistics of interest \cite{61_Varian, 62_Weisstein}. We used the results from our bootstrapping to estimate our key statistics and provide their confidence intervals. 

\input{content_minor_revision__Apr2022/figure/bootstrap_workflow_2}

\subsection{\rnr{Bootstrapping Resampling Process}}
We took the following steps to sample our data for bootstrapping. For simplicity, we will describe our method with reference to the May 2016--March 2017 dataset, which we will refer to as $T_{2016}$. 

Intuitively, a bootstrap uses resampling with replacement to create a large number of parallel universes, each with the same number of comments as the original dataset, but each universe will have a slightly different number of norm violating comments due to the resampling with replacement. This variation across many parallel universes is what yields uncertainty confidence intervals on our estimates. However, each universe (bootstrap sample) must also contend with the fact that we have not manually annotated all 5 million comments in the dataset as violating or not. Rather than use a single fixed measurement of norm-violating comments per subreddit, which would ignore this source of uncertainty in the estimation, we resample our annotations over and over again with replacement to build in uncertainty due to our limited number of annotations. Figure~\ref{fig:bootstrap} provides an overview of this process, which is explained in detail below.

\subsubsection{\rnr{Study data and the flagged comments}} 
We began by randomly sampling a set of unmoderated comments. Specifically, we sampled 5,000 random comments from each of the study subreddits posted during this period dataset, for a total of 485,000 comments. These comments were sampled from the Pushshift Reddit corpus, which contains a complete capture of all comments on each subreddit. For clarity in presentation, we will call this sample of comments from 2016--2017, $T_{2016}$. We then ran all 97 subreddit classifiers on each of the sampled comments to calculate the classifier agreement score. Any comments that were flagged by at least 80 of the 97 classifiers as violating were labeled as \textit{flagged}.

\subsubsection{\rnr{Calculating intermediate probabilities}}
For the bootstrap, we needed to follow a procedure where we sample a comment, label it as flagged or not based on the classifiers, and then label flagged comments as violating or not based on human annotators. However, because we cannot tractably label every flagged comment in the bootstrap via human annotation, we relied on statistical generalization via the bootstrap. For this generalization to succeed, we require two empirically observed probabilities for each bootstrap iteration: $P_{subreddit}(violating \mid flagged)$, the true positive rate, the probability that a flagged comment in a particular subreddit is verified by annotators as an actual violation; $P(violating \mid \lnot flagged)$, the false negative rate, the probability that a comment that is not flagged as potentially violating via our classifiers is in fact violating according to our annotators.

\paragraph{$P_{subreddit}(violating \mid flagged)$.} The true positive rate, $P_{subreddit}(violating \mid flagged)$, carries uncertainty since we can only manually annotate a fixed number of comments per subreddit. To model this uncertainty, we re-estimate $P_{subreddit}(violating \mid flagged)$ with each bootstrap resample. Each iteration, for each subreddit, we take a random resample with replacement of the 32 flagged comments that annotators had labeled for each subreddit with each resample (e.g., sample 32 comments with replacement from the set of 32 flagged comments for the subreddit). This value of $P_{subreddit}(violating \mid flagged)$ is then used for that subreddit for that bootstrap iteration and varies with each iteration, capturing uncertainty in our estimate, and is used to help estimate whether a flagged comment in our sample is a true positive.

\paragraph{$P(violating \mid \lnot flagged)$.} Some comments that the classifiers do not believe to be macro norm violations will, in fact, be violations---and our bootstrap procedure must account for this so that it does not undercount the number of violations. To estimate the false negative rate, $P(violating \mid \lnot flagged)$, we use the sample of 1,000 \textit{non-flagged} unmoderated comments that we manually annotated in Section 3.4. These 1,000 comments showed a consistent 1\% false negative rate of our classifiers across 10 randomly chosen subreddits, so we treat this value as being consistent across all subreddits and do not parameterize it by subreddit. However, we still must model uncertainty in this estimate. So, for each bootstrap iteration, we resample 1,000 comments with replacement from the set of 1,000 comments and calculate $P(violating \mid \lnot flagged)$ as the proportion of comments that were not flagged as potentially violating but were in fact violating according to human annotators. We use this quantity to estimate whether a non-flagged comment in our sample is a false negative. As with the true positive rate, resampling with replacement will cause this value to vary from bootstrap iteration to iteration, critical to creating confidence intervals for our analysis.

\subsubsection{\rnr{Bootstrap resampling iteration procedure}}
These two probabilities enable our bootstrap procedure. In bootstrapping, we resample the dataset many times and measure the quantity of interest in each new instance. \rnr{In this study, we ran our bootstrapping procedure for 1,000 iterations (this is a typical number of iterations when conducting bootstrap resampling \cite{81_Bootstrap}). For each iteration, we resampled a number of comments that matched the complete number of comments on each of our study subreddits during our study period (i.e., 252,642,908 comments for $T_{2016}$). We also resample with replacement from the dataset of 5,000 classified comments each iteration, creating variation to model uncertainty in our sampling. Each bootstrap iteration also estimates new values for $P_{subreddit}(violating \mid flagged)$ and $P(violating \mid \lnot flagged)$ for each iteration, which combine to provide one datapoint in the final outcome distribution. In other words, each bootstrap sample creates an alternative universe of comments resampled from each subreddit, exhibiting natural variation in violation and moderation rates due to the random sampling.}

We illustrate this process with an example of bootstrap resampling the r/videos subreddit for $T_{2016}$. There were a total of 5,296,900 unmoderated comments that were on r/videos during $T_{2016}$. To populate each of these comments, we sampled a random comment from the subset of 5,000 random comments from r/videos that the classifiers had labeled. These 5,000 random comments had themselves been randomly sampled with replacement for this iteration of the bootstrap from the original set of 5,000 comments that were collected and classified. For r/videos, in one of the bootstrap samples, 1,249 of the 5,000 randomly sampled unmoderated comments (25\%) were flagged as potentially violating by the classifiers. If the sampled comment was flagged, then we randomly assign the comment as violating with probability $P_{r/videos}(violating \mid flagged)$ ($\approx 0.2$ in this bootstrap iteration) and attach the violation type(s) that human annotators tagged that comment with, and as not violating otherwise. If the sampled comment was not flagged, then we randomly assign the comment as violating with $P(violating \mid \lnot flagged)$ ($\approx 0.01$ in this iteration), and not violating otherwise. This process repeats to generate all 5,296,900 comments for r/videos.

We follow this process for all 97 subreddits, for all 1,000 bootstrap samples. The 1,000 samples provide a distribution and uncertainty estimate for the key statistics. Finally, through this process, we resample the same number of comments as the number of all unmoderated comments that were posted on these subreddits, some of which are labeled to be violating. So when we calculate the rate of unmoderated but violating comments in one bootstrap iteration, we add up the number of unmoderated but violating comments in each of the 97 subreddits and divide that by the number of all comments --- this gave us an estimation for r/videos, which was 5.95\% [2.16, 9.22] for $T_{2016}$, and 3.53\% [0.59, 9.14] for $T_{2020}$. This allows our bootstrap procedure to naturally take into account the different subreddit sizes --- through resampling more comments from the larger subreddits, the bootstrap will give more weight to the uncertainty on larger subreddits, in the analysis which are focused on the overall rates across the platform.

\subsubsection{2020 dataset}
As mentioned in the previous section, there are two periods of interest: one, from May 2016 to March 2017, the same period as when the moderated comments dataset $\mathcal{M}$ was collected, and the other from the last three weeks of December 2020, which we consider as a replication study on a slightly shorter but more recent timeframe. As we referred to the original dataset as $T_{2016}$, we will refer to the latter as $T_{2020}$. The process for $T_{2020}$ was identical to $T_{2016}$. Because $T_{2020}$ was a smaller dataset, however, its constituent samples were fewer in number. In particular, we \rnr{randomly} sampled 2,000 comments from each of the study subreddits instead of 5,000 commments, or fewer if the subreddit did not contain 2,000 comments during $T_{2020}$. This resulted in a total of 188,000 comments sampled and run through our 97 subreddit classifiers. Given the smaller sample size, the confidence intervals are wider for $T_{2020}$ than for $T_{2016}$.

\subsection{\rnr{Ablation Analysis with Fewer Annotations}}
Our bootstrap relies on a set of thousands of manually annotated comments from annotators. This gives rise to a possible threat to validity: that we did not manually annotate enough comments per subreddit to capture the uncertainty in the estimate per subreddit. To test whether this threat should be concerning, we performed an ablation analysis where we replicated our method using half of the manual annotations we had for each of the subreddits (that is, 16 annotations per subreddit instead of 32 for $T_{2016}$, and 4 per subreddit instead of 8 for $T_{2020}$). The main goal of this ablation analysis was to test how having fewer annotations impacts the uncertainty estimates of the bootstrap. Intuitively, as we have fewer and fewer samples, the confidence intervals will widen because if a rare event is sampled, it has a larger impact on the resulting estimate. We report our findings alongside our main results, in Section 6.

%% file: content_minor_revision__Apr2022/figure/bootstrap_workflow_2.tex
\begin{figure}[tb]
  \centering
  \includegraphics[width=1.0\textwidth]{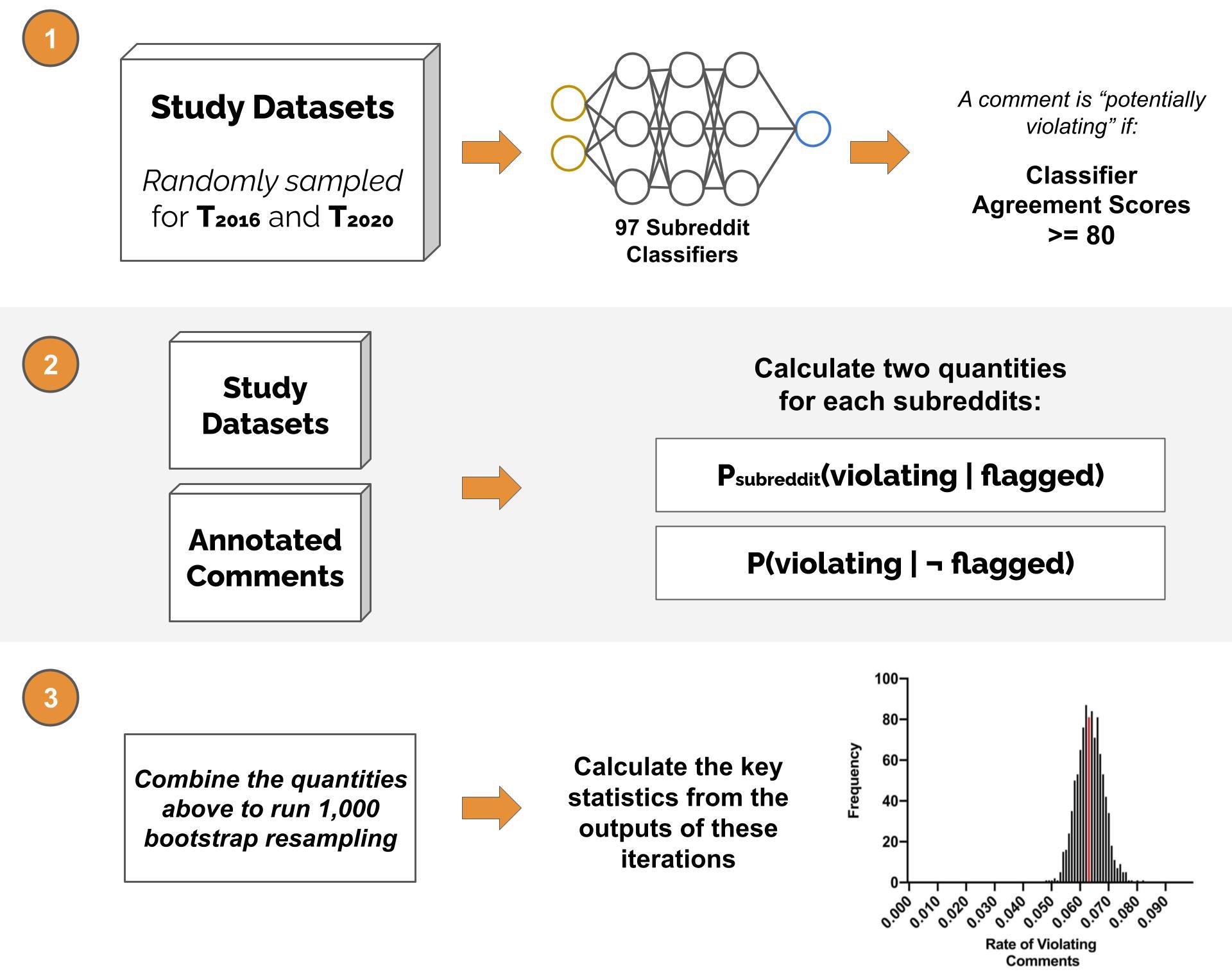}
  \caption{\rnr{An overview of the sampling process in broad strokes. We start by randomly sampling unmoderated comments from $T_{2016}$ and $T_{2020}$ to generate our study datasets. We use these datasets and the annotated comments to calculate key quantities needed for our bootstrap resampling. We then combine these quantities to implement our resampling procedure, which we repeat 1,000 times for each time periods of our interest.}}
  \label{fig:bootstrap}
  \Description{Bootstrap resampling workflow}
\end{figure}

%% file: content_minor_revision__Apr2022/method-pt3.tex
\section{Measurements}
Using our pipeline for identifying violating comments and bootstrap sampling of comments, we estimated the dependent variables of interest. In this section, we present our specific measures that we derived from our method.

\subsection{Estimating the Prevalence of Violating Comments}

\subsubsection{Overall prevalence} 
Our core dependent variable is the percentage of unmoderated comments that are \rnr{macro norm} violations. For each bootstrap sample, we calculated the percentage of violating comments. We then calculated our estimate as the median result across the 1,000 bootstrap samples, and calculated the 95\% confidence interval across those samples. 

\subsubsection{Prevalence per subreddit} 
Do some subreddits contain a higher rate of macro norm violating comments than the others, and if so, why? To answer this question, we estimated the percentage of violating comments for each of the subreddits across each of the bootstrap samples, and again calculated the median and 95\% confidence interval per subreddit. Given that the $T_{2020}$ sample, as mentioned previously, is a smaller scale replication of the $T_{2016}$ sample, we found that there is not enough data to draw statistically meaningful results for each of the subreddits for $T_{2020}$. Therefore, we focused our analysis in this particular subsection on the $T_{2016}$ sample. 

We tested the factors that were associated with a subreddit having a higher count of violations. We started by taking the sampling approach as we did above to estimate the count of violating comments for each of the study subreddits. We hypothesized that two relevant predictive variables might influence the rate of violating comments: 1) the topic of a given subreddit and 2) the ratio of the number of moderators to the number of comments. Our focus on the first variable was motivated by our observation that some communities might be more prone to hosting macro norm violations that are topically relevant (e.g., politically inflammatory comments for political subreddits or pornography for NSFW subreddits), while our focus on the second variable was informed by prior work suggesting that moderators in subreddit communities are overloaded and unable to review every comment posted~\cite{Chandrasekharan2018internet, 18_Gilbert}.

To test these hypotheses, we first collected the relevant data for these variables. To get the respective topics for our study subreddits, we matched the study subreddits to categories on r/ListOfSubreddits’ thematically organized list of subreddits.\footnote{\url{https://www.reddit.com/r/ListOfSubreddits/wiki/listofsubreddits}} The list contains multiple layers of thematic hierarchy. From this, we picked the second-highest thematic layer (e.g. r/AskReddit was categorized as “Discussion,” and r/Pokemon as “Entertainment”) for our subreddits as the top layer was too broad to be meaningful. This resulted in eleven mutually exclusive topic categories for our 97 study subreddits.  We then manually collected the number of moderators who were present in October 2016 (the middle of the 11-month period when $\mathcal{M}$ was collected) for each of the subreddits by using the Wayback Machine to access an archived version of each subreddit's homepage on October 1st, 2016, or the earliest subsequent date when the page was crawled. We then calculated the ratio of moderators to comments for each of the subreddits. Finally, we retrieved the number of all comments (including those that were removed by the users or moderators) that were posted on each of the 97 study subreddits from $T_{2016}$ -- we used this information as the offset to our Poisson model described below. To address the non-normal distribution of the number of comments and moderator-to-comment ratio in our regression, we log transformed these variables.

We then employed a Poisson regression \cite{63_poisson} to model the rate of macro norm violating comments in a given subreddit using the three aforementioned predictive variables as our independent variables. \rnr{Poisson regression was an appropriate choice because it is designed to model rate data. We followed a common practice of setting an offset---in our case, the size of the subreddit in terms of the number of comments--- to the Poisson regression model to make it appropriate for rate data \cite{94_Anderson, 95_Gardner}.} For better interpretability of our model, we treated the ``general content'' topic category as the baseline of the indicator (dummy) variable for topic. Finally, to interpret our results, we exponentiated the variable coefficients in our model to calculate the incidence rate ratios. For instance, as we will cover in the next section, \rnr{the number of moderators to comments ratio in a subreddit is a significant predictor of the rate of violating comments in a subreddit (p<0.001) where for every unit increase in the log-transformed number of comments in a subreddit, the rate of violating comments is expected to increase by a factor of 2.07 (=$e^{0.728}$).}

\subsection{Determining the Characteristics of Violating Comments}

\subsubsection{Prevalence by violation category} 
So far, we defined a comment to be violating if it violates one of the macro norms of Reddit. But here, we describe our analysis that takes a closer look at the variance in the prevalence of different macro norm violations and the rate at which they occur. In order to better understand the prevalence of each of the macro norm violations among the unmoderated comments, \rnr{we measure the percentage of sampled violations that fall into each of the violation categories}. Specifically, we take the average number of norm violations per category that across the study subreddits in each bootstrap iteration, and as before, report the median and confidence interval across the 1,000 bootstrap iterations.

Are all categories of macro norm violations removed by the moderators at the same rate? To answer this question, we \rnr{take the following steps to} estimate the number of all comments (moderated and those still online) that violate each specific one of the macro norms. To estimate the number of violations for each of the macro norms within the moderated comments, we require a complete set of all moderated comments from our study subreddits, which we gathered earlier as $\mathcal{M}$ for $T_{2016}$. To extend $\mathcal{M}$ to $T_{2020}$, we reimplemented the original high-frequency scraping process used in prior work to generate $\mathcal{M}$ for $T_{2016}$~\cite{Chandrasekharan2018internet}. We then randomly sampled from $\mathcal{M}$ (N=776 comments for the longer $T_{2016}$ period; N=188 for the shorter $T_{2020}$ period). We then followed the same human annotation process from Section~\ref{sec:annotation} to annotate each moderated comment with any macro norms that it violated. Using this, we calculated the estimated number of comments that violate each of the macro norms within the moderated comments dataset. We combined this with the estimated number of unmoderated violations for each macro norm to estimate the overall number of violations for each of the macro norms considering both moderated and unmoderated comments combined. With this number of violations, we then proceeded to calculate the rate at which each of the macro norms are moderated.

\subsubsection{Comparing the rate of engagement} 
Reddit comments accrue engagement over time in the form of upvotes, downvotes, and replies. The “score” of a comment is determined by subtracting the number of downvotes from the number of upvotes the comment gained over time, and “replies” are comments that respond to the given comment in the thread. Score is a useful dependent variable to track because it is a strong determinant of how high in the thread the comment sits.\footnote {On Reddit, users can choose one of the three options -- top, hot, and best -- for determining what comments should come at the top. Top is simply whichever comment has the highest score, hot is the log-transformed absolute value of the score with extra weights for the age of the content, and best is the number of upvotes divided by the number of downvotes. See \url{https://www.reddit.com/r/explainlikeimfive/comments/1u0q4s/eli5_difference_between_best_hot_and_top_on_reddit/}} The number of top-level replies can serve as a proxy for measuring whether the given comment generated, or ended, a discussion.\footnote{The Perspective model from Jigsaw for measuring toxicity used human annotations to train where a comment was asked to be annotated if it were “rude, disrespectful, or unreasonable… that is likely to make you leave a discussion” \cite{60_Perspective}.}
We explored whether the violating comments result in significantly higher or lower scores, and whether they receive more or fewer “top-level replies” (replies that directly respond to the given comment in the thread of replies). To investigate this, we compared the scores and the number of top-level replies that the unmoderated violating comments got to the global average number of online comments the study subreddits in our $T_{2016}$ and $T_{2020}$ samples.

\subsubsection{Comparing language usage} 
Does the way in which a macro norm is violated impact its likelihood of remaining on the site or being moderated? We compared the writing style of moderated comments vs. unmoderated violating comments on two dimensions: readability and emotionality. For measuring readability of a comment, we used the Flesch–Kincaid readability test~\cite{64-kinkaid} to retrieve the comment’s readability score, where a higher score indicates that the comment is easier to read. For measuring emotionality of a comment, we used Evaluative Lexicon 2.0~\cite{65_Rocklage, 66_Rocklage}, an imputation-based dictionary that assigns an emotionality score to text where higher score means that the comment is more likely to trigger an emotional reaction rather than a cognitive response that reflects a person’s beliefs about the topic of discussion \cite{66_Rocklage}.

%% file: content_minor_revision__Apr2022/results.tex
\input{content_minor_revision__Apr2022/figure/bootstrap_histogram} 
\input{content_minor_revision__Apr2022/figure/all_rate_of_violating}

\section{RESULTS}

In this section, we present the results of our analyses. We start by reporting on the prevalence of violating comments across our study subreddits and then describe their characteristics in terms of their content, rate of engagement, and language usage.

\subsection{Prevalence of Violating Comments}

\subsubsection{Macro norm violating comments are common}
Figure~\ref{fig:bootstraprate} reports the rate of macronorm violating comments across our 1,000 bootstrap iterations. In $T_{2016}$, 6.25\% (95\% CI [5.36\%, 7.13\%]) of all unmoderated comments are violations. In $T_{2020}$, 4.28\% (95\% CI [2.50\%, 6.26\%]) of all unmoderated comments are violations. \rnr{There are fewer macro norm violating comments over time: a permutation test confirms that the rate of macro norm violating comments is lower in $T_{2020}$ than $T_{2016}$ ($p<.001$).} 

\subsubsection{\rnr{Ablation analysis of the bootstrap confidence intervals}} 
\rnr{In order to (1) confirm that our bootstrapping procedure widens its confidence intervals as fewer human annotations are available, as it should, and (2) confirm that our current number of annotations is sufficient for our estimation goal, we tested the width of the bootstrap's confidence intervals as we ablated to half the number of human annotations per subreddit. On our full dataset for $T_{2016}$ with 32 human annotations per subreddit, the 95\% confidence interval's width is 1.77\%. When we ablate and halve the number of labels from 32 to 16 per subreddit, the confidence interval's width becomes $T_{2016}$ becomes  2.58\% ([4.60, 7.18], median 5.90\%), a 95\% CI width that is .8\% larger. For our smaller $T_{2020}$ dataset with 8 annotations per subreddit, the 95\% CI width is 3.76\% ([2.50, 6.26]). Here, when we ablate and halve the number of labels from 8 to 4 per subreddit, the 95\% CI width widens to 5.43\% ([1.96, 7.39], median 4.27\%), or 1.67\% wider than the original estimate.}

\input{content_minor_revision__Apr2022/table/poisson_results}

\rnr{Two observations fall out of this ablation analysis. First, we observe that the width of the confidence interval widens as we have fewer annotated comments per subreddit: from 1.77\% with 32 annotations to 2.58\% with 16 annotations in 2016, and from 3.76\% with 8 annotations to 5.43\% with 4 annotations in 2020. This confirms that bootstrapping is not producing overconfident confidence intervals: the confidence intervals widen as we have fewer and fewer labels. Second, we observe that, for 2016 in particular, 32 labels is far more than necessary for the confidence interval to be tight: the interval does not begin to dissolve until having only four labels per subreddit, and providing more than 32 labels per subreddit will not substantially reduce the width of the confidence interval further. The reason for this stability is that, while the estimate for any particular subreddit may be more uncertain, aggregating over nearly 100 subreddits substantially reduces the overall uncertainty.}

\subsubsection{\rnr{Topic and moderator ratio predict the rate of macronorm violations}}
In this subsection, we focus our efforts on the $T_{2016}$ sample. The rate of \rnr{macro norm} violating comments for individual subreddits averages 4.9\% (std=4.0\%), but there is a large spread across different subreddits (Figure 5). For example, subreddits such as r/AskHistorians (0.90\%; 95\% CI [0.73\%, 1.12\%]) and r/science (1.10\%; 95\% CI [0.70\%, 1.61\%]) had lower rates of macro norm violating comments than ones such as r/atheism (10.06\%; 95\% CI [6.99\%, 13.27\%]) and r/politics (14.46\%; 95\% CI [10.79\%, 18.12\%]). Our Poisson regression model, summarized in Table 3, suggests that these differences are associated with the subreddit's topic and its moderator to comment ratio. More specifically, subreddits on NSFW (p<0.001) or political (p=0.016) topics have higher a rate of violating comments, and those hosting hobbies and occupation topics (p=0.001), or discussion topics (p=0.018) are associated with fewer violating comments. Additionally, higher moderator-to-comment ratios are associated with fewer violating comments (p=0.003). 

Although NSFW and political topics featured a higher rate of macro norm violating comments, this did not mean that subreddits covering topics other than these two necessarily had a small number of macro norm violating comments. When excluding 17 subreddits whose topics were categorized as either NSFW and political, we find that the remaining subreddits still had violating comments at the rate of 5.26\% (95\% CI [4.27\%, 6.39\%]).

\input{content_minor_revision__Apr2022/figure/prevalence_by_mn_categories}

\subsection{Characteristics of Violating Comments}
In this subsection, we delve deeper into the violating comments to explicate the categories of violation, engagement rate, and language usage.

\subsubsection{Personal attacks are the most prevalent category of violation.} 
The eight macronorms are not equally prevalent (Figure 6). In $T_{2016}$, personal attacks (48.58\%; 95\% CI [40.66\%, 56.66\%]) and politically inflammatory comments (30.16\%; 95\% CI [22.80\%, 37.54\%]) are particularly prevalent macronorm violations. On the other hand, categories such as posting pornographic links (<1\%) or abusing moderators (<1\%) were less common. Similarly, in $T_{2020}$, personal attacks (61.94\%; 95\% CI [45.06\%, 80.40\%]) remain the most prevalent, misogyny/vulgarity violations have become more prevalent (20.90\%; 95\% CI [1.33\%, 40.67\%]), and politically inflammatory comments have become less common (14.19\%; 95\% CI [4.49\%, 30.37\%]).

\subsubsection{Macronorms are moderated at different rates} Moderation removed 4.86\% of macronorm violating comments in 2016–2017, and 10.54\% in 2020. Not all types of macro norm violations were moderated at the same rate, however.
In $T_{2016}$, comments that include links to pornography were moderated at the highest rate (34.53\%), followed by those that abuse or criticize moderators (26.17\%) and those that accuse another of being sensitive (12.87\%). On the other hand, politically inflammatory comments (4.30\%) and misogyny/vulgarity (4.13\%) were less likely to be moderated. Some trends still held in $T_{2020}$, where comments that include pornography were moderated at the highest rate (45.04\%). Nearly all categories were more heavily moderated in 2020 than in 2016.
We summarize the results in Figure 7.

\subsubsection{Violating comments get fewer upvotes}
We find that in $T_{2016}$, \rnr{macro norm} violating comments garnered an average score of 7.32 (std=21.97), which is significantly lower than the global average of 10.27 (std=95.95) in our dataset according to Welch's t-test (t(469647) = -3.29, p = .001). However, there was no significant difference in the number of top-level replies; the violating comments got 2.00 top-level replies (std=9.25), vs. the global average of 1.92 (std=7.32): t(469647) = 0.21, p=0.83. These result replicate in the $T_{2020}$ sample; the comments that were annotated as violating got an average of 6.07 upvotes (std=13.09), which is significantly lower than the global average of 12.17 (std=119.94) that our sample of 188,000 comments got (t(415515) = -4.30, p < .001). Similarly, there was no significant difference in the number of top-level replies; the violating comments got 1.77 top-level replies (std=2.81) where  1.77 (std=6.82) was the global average (t(415515) = -0.003, p=0.99).

\input{content_minor_revision__Apr2022/figure/rate_of_moderation_by_mn}

\subsubsection{More readable and emotional comments are more likely to remain unmoderated} We find that in $T_{2016}$, the readability scores of unmoderated macro norm violating comments averages 62.38 (std=122.51) while the scores of moderated comments averages 33.59 (std=1849.25). Additionally, using Welch's t-test, we confirm that the difference in readability between these two groups of comments is significant (t(2007616)=5.62, p<0.001), indicating that the unmoderated macro norm violating comments are easier to read than the moderated comments. This is replicated in $T_{2020}$, in which the readability scores of the unmoderated macro norm violating comments averages 76.95 (std=31.94) while the scores of moderated comments averages 64.35 (std=293.49). Additionally, this difference between the two groups remains significant (t(160481)=3.59, p<0.001). 

Similarly, we find that in $T_{2016}$, the average emotionality scores of unmoderated macro norm violating comments averages 5.14 (std=1.30) while those of moderated comments averages 0.74 (std=0.51). Once again, the difference in terms of the emotionality score between these two groups of comments is significant according to Welch's t-test (t(378432)=59.89, p<0.001), indicating that the unmoderated macro norm violating comments are more emotional than the moderated comments. This is replicated in $T_{2020}$ in which the average emotionality scores of unmoderated macro norm violating comments averages 4.75 (std=1.40) while those of moderated comments averages 0.80 (std=0.49) with the difference between the two groups being significant (t(74126)=40.90, p<0.001).

%% file: content_minor_revision__Apr2022/figure/bootstrap_histogram.tex
\begin{figure}[tb]
  \centering
  \includegraphics[width=0.95\textwidth]{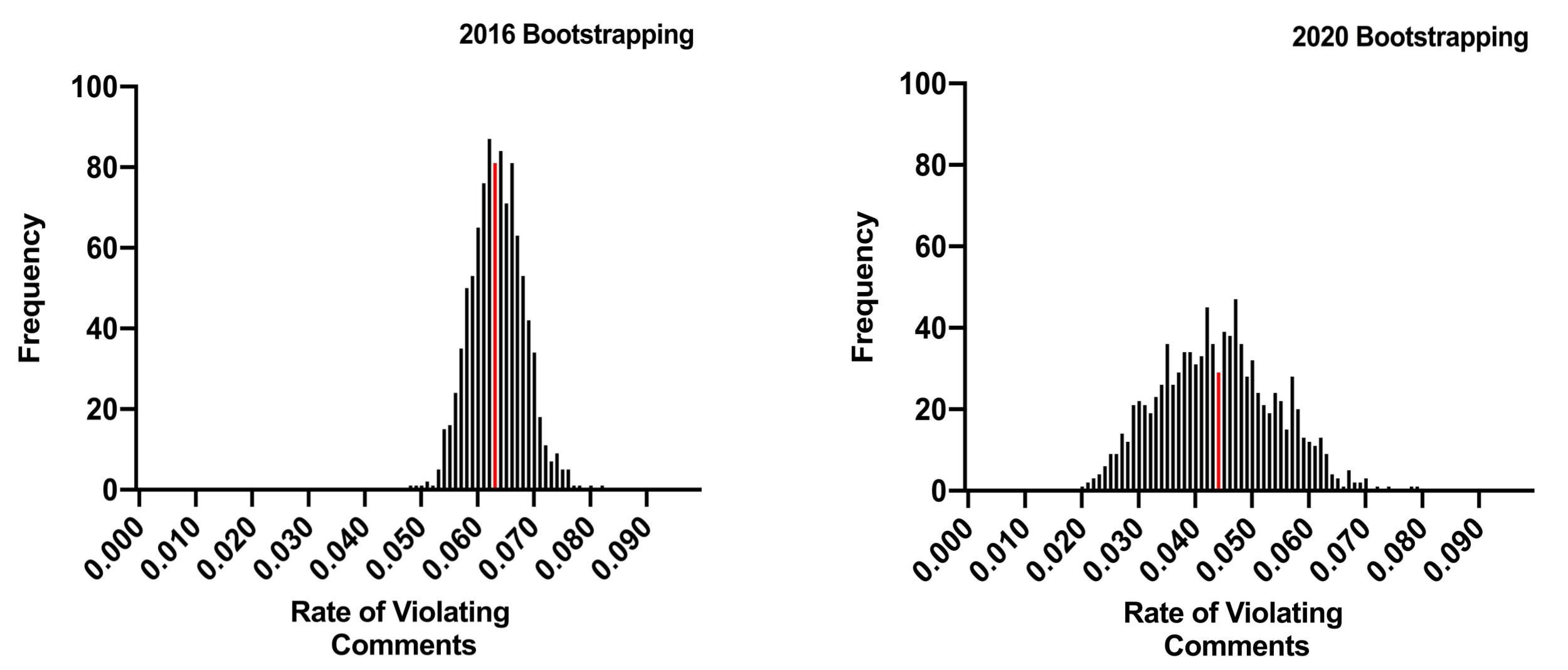}
  \caption{The results of our bootstrap sampling. The figures show the overall rates of violating comments that are left unmoderated for all 1,000 simulations of Reddit for $T_{2016}$ and $T_{2020}$. The 50th percentile rates for the 1,000 bootstrap samples are colored red for the two periods.}
  \label{fig:bootstraprate}
  \Description{Annotation Interface}
\end{figure}

%% file: content_minor_revision__Apr2022/figure/all_rate_of_violating.tex
\begin{figure}[tb]
  \centering
  \includegraphics[width=0.95\textwidth]{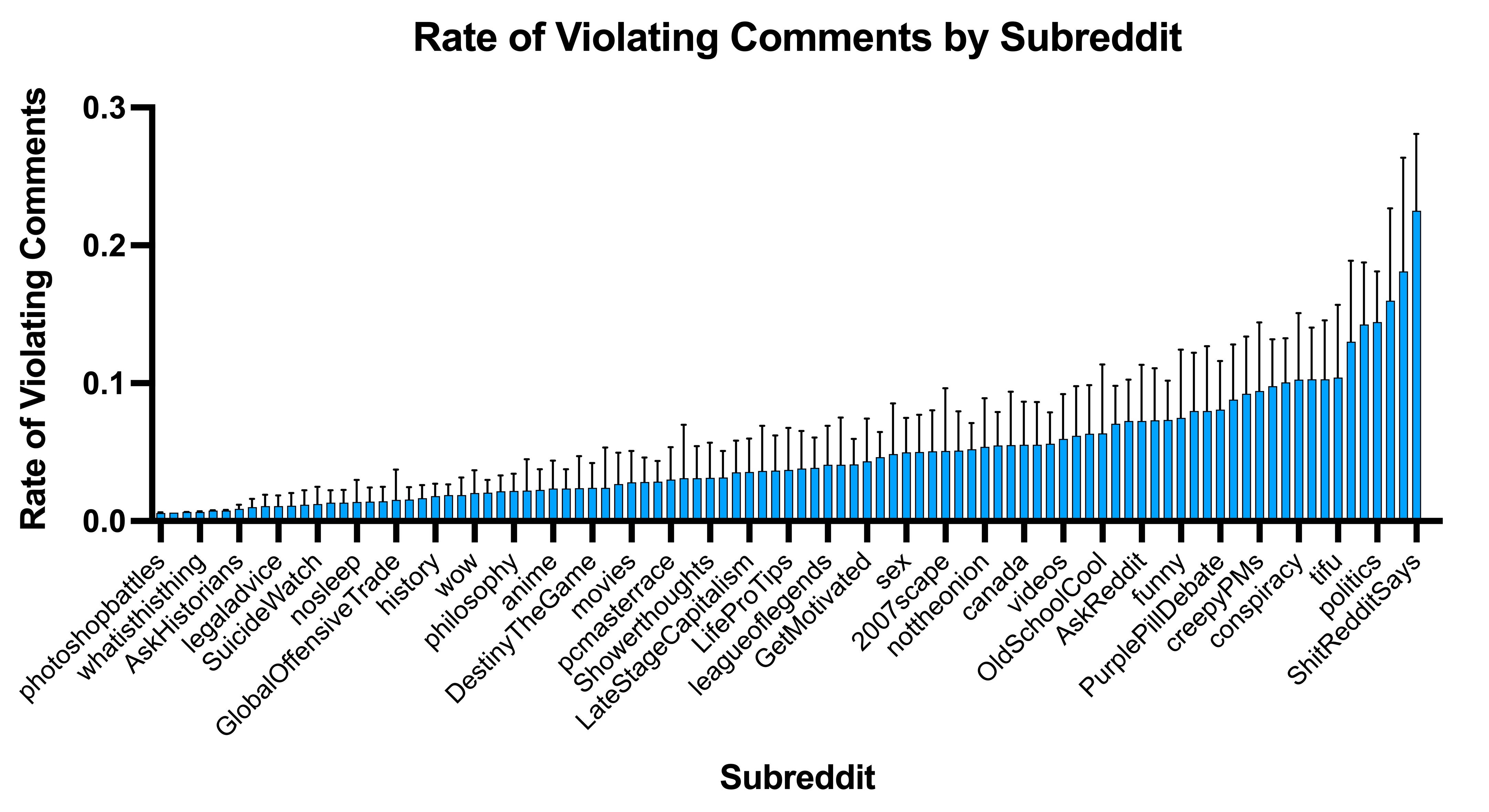}
  \caption{Rate of \rnr{macro norm} violating comments by subreddit in $T_{2016}$. The lighter lines indicate 95\% confidence intervals. Subreddits vary in levels of violating content; some subreddits like r/depression (1.10\%) and r/Android (1.42\%) contain almost no  \rnr{violating} comments online, whereas some like r/creepyPMs (9.42\%) and r/EnoughTrumpSpam (14.27\%) contains much higher rates of \rnr{violating} comments.}
  \Description{Rate of macro norm violating comments by subreddits}
\end{figure}

%% file: content_minor_revision__Apr2022/table/poisson_results.tex
\begin{table}[tb]
\centering
\captionof{table}{The results of Poisson regression that predicts the rate of problematic comments in a subreddit. We find that the number of comments in a subreddit, as well as a few of the topic categories are strong predictors of the rate of problematic comments a subreddit may have. The topic "general content", which was included in the model, is not shown as it was used as the baseline dummy variable for the categorical variable.}
\begin{tabular}{lllll} 
\toprule
                               & \textbf{Coefficient} & \textbf{SE} & \textbf{z-score} & \textbf{p-value}  \\ 
\toprule
Intercept                      & -5.919               & 1.249       & -4.740           & < 0.001 (***)      \\
log(mod to comment ratio)      & -0.728               & 0.244       & -2.986           & 0.003 (**)        \\
Topic: Hobbies and Occupations & -2.048               & 0.280       & -7.304           & < 0.001 (***)      \\
Topic: NSFW                    & 1.958                & 0.253       & 7.742            & < 0.001 (***)      \\
Topic: Discussion              & -1.021               & 0.421       & -2.423           & 0.015 (*)         \\
Topic: Politics                & 0.628                & 0.260       & 2.414            & 0.016 (*)         \\
Topic: Humor                   & 0.408                & 0.251       & 1.621            & 0.105             \\
Topic: Lifestyle               & 0.0879               & 0.321       & 0.271            & 0.787             \\
Topic: Educational             & -1.191               & 0.647       & -1.840           & 0.078             \\
Topic: Entertainment           & -0.165               & 0.285       & -0.579           & 0.563             \\
Topic: Technology              & -0.387               & 0.325       & -1.191           & 0.233             \\
Topic: Other                   & 0.692                & 0.324       & 2.135            & 0.033             \\
\bottomrule
\end{tabular}
\end{table}

%% file: content_minor_revision__Apr2022/figure/prevalence_by_mn_categories.tex
\begin{figure}[tb]
  \centering
  \includegraphics[width=1\textwidth]{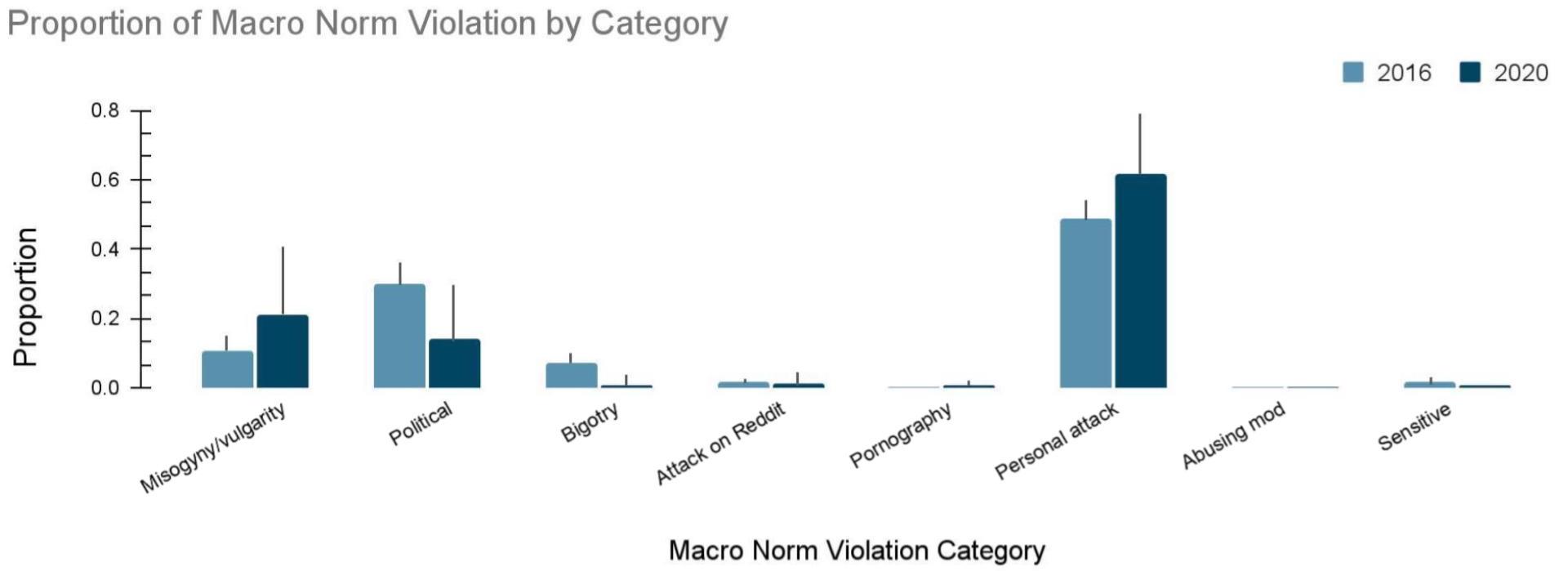}
  \caption{The proportion of macro norm violation by the eight categories among the unmoderated problematic comments according to our annotators. In $T_{2016}$ as well as $T_{2020}$, personal attacks were the most common macro norm violations. Politically inflammatory comments, on the other hand, have become less prevalent.}
  \Description{Proportion of Macro Norm Violation by Categories}
\end{figure}

%% file: content_minor_revision__Apr2022/figure/rate_of_moderation_by_mn.tex
\begin{figure}[tb]
  \centering
  \includegraphics[width=1\textwidth]{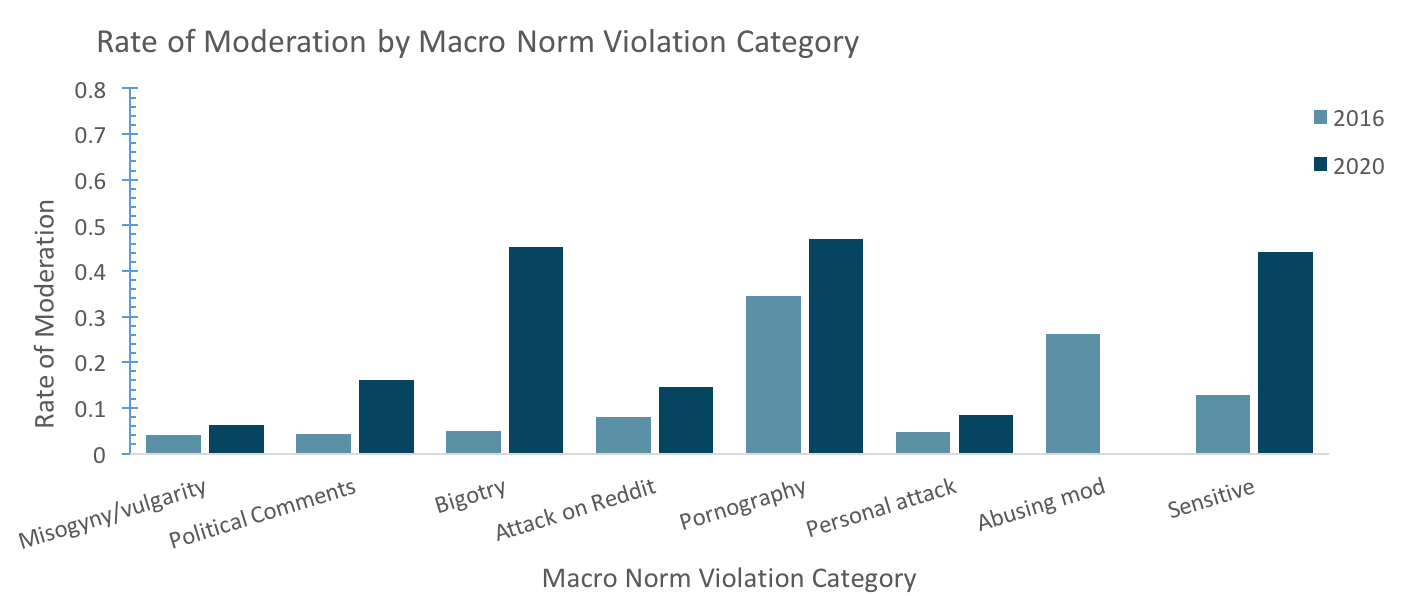}
  \caption{The rate of moderation by the eight macro norm violation categories acquired through stratified sampling. In $T_{2016}$, comments that contain links to pornography and comments that abuse or criticize moderators were moderated at the highest rate. In $T_{2020}$, comments with links to pornography is still among the most commonly moderated categories of norm violation. In addition, the overall rate of moderation has increased.  }
  \Description{Rate of moderation by macro norm violation categories}
\end{figure}

%% file: content_minor_revision__Apr2022/discussion.tex
\section{Discussion}
The contribution we make in this paper is twofold. First, we present a large-scale analysis that investigates the prevalence and characteristics of macro norm violating comments that are left unmoderated on Reddit. In doing so, we provide the social computing literature with empirical results that can help aid policies, moderation, and design interventions. These results illustrate how roughly one in every 20 comments on Reddit violates the platform's own expectations of what should not be present, and shines light on the the most prevalent types of violations. Second, we demonstrate a materialized human-AI-bootstrap pipeline for identifying violating comments. We use classifiers trained on a dataset of previously moderated comments on the 97 most popular subreddit communities on Reddit to triage potentially violating comments at scale, employ trained human annotators to validate a sample of the classifiers’ decisions, and calculate the final estimation and our confidence using the bootstrap resampling technique.

In the remainder of this section, we discuss what our findings may mean for social media and for efforts to rein in anti-social behavior online; we try to interpret what the rate of violating comments in the 4\% to 6\% range means in reality, and scope out what can be inferred from our results. Finally, we end with our work’s limitations and potential avenues for future research.

\subsection{Is This Result Good, Bad, or Neither?}
Comments that violated the macro norms were prevalent on the study subreddits during both time periods. As noted in the introduction, even Facebook's transparency report of 0.1\% of its content being categorized as hate speech~\cite{58_Culliford, 91_Barrett} raised concerns as it translates to affecting millions of users. Given this, 4\% to 6\% of norm violating comments likely is not a small amount. Also, the overall rate of moderation, though has increased over the years perhaps in some part due to the growing awareness of the negative impacts these content can have, remains low with one in twenty of the norm violating comments in 2016, and one in ten in 2020, being removed by the moderators. These rates are comparable to the 3--5\% rate of hate speech moderation reported in Facebook’s internal documents that were made public in recent whistleblower exposures \cite{facebook_scandal, facebook_scandal_2}. This provides an important context for both our findings and those in Facebook’s internal documents; that these two findings covering very different contexts and likely, using diverging methods of measurement (it is unclear how Facebook measured their moderation rate), are roughly converging may illustrate a broader point about the challenges of today’s large-scale content moderation.

\mrev{It is also worth highlighting that for many communities we studied in this work, our results on the prevalence of norm violating comments likely represents the lower bound for two reasons. First, we are explicitly not measuring the prevalence of the comments that do not violate the macro norms, yet still violate community-specific local, micro norms. For example, subreddits for minority groups often have strong prohibitions on attacks, and topic-based groups such as r/science remove all comments except those focusing on the science itself. Second, our approach for identifying violating comments was based on reviewing the comment in isolation and therefore did not include norm-violating comments that were violating only in a specific conversational context.} These factors further highlight that despite the growing effort to moderate violating content online, content moderation still remains challenging with much work to be done. 

\subsubsection{\mrev{What Does Not Get Moderated?}}
Our findings also suggest that there are certain categories of norm violations and community specific characteristics that make it more or less likely for violating comments to be moderated on Reddit. For instance, comments that contain pornographic materials were moderated almost at the rate of nearly 50\% in 2020. Of course, not all violations are created equal: some categories are more harmful than others, and violations may be more or less harmful based on who they're directed to and in what context. Our analysis cannot measure this; it is more effective at identifying the overall volume of microaggressions and micro-violations. However, by combining our approach with survey analyses, it might be possible to weigh the impact of each violation to produce a better overall sense of which groups are harmed, when, and how. We also note that higher rates of moderation for some categories may be due to platform incentives (e.g., will Reddit threaten the subcommunity if they are not removed), as well as the ease of automatic moderation. Pornography, for example, may be more readily identifiable via URL by regex-based AutoModerators on Reddit.

\subsection{\mrev{What Should the Moderation Community Aim For?}} 
\mrev{The estimation we present here may seem like cause for concern---many macro norm violating comments in these online communities remain unmoderated. The important question then is, where do we go from here? What is a reasonable goal the moderation community should strive for? We believe that it would be a mistake to take our results presented here to mean that the moderation community is failing or needs to put in even more work going forward. Particularly in the context of Reddit, it is important to note that communities are largely self-governed and the moderation effort falls singularly on unpaid, and often overworked, volunteers~\cite{Kiene2019, Gilbert2020}, so the amount of available labor might be fundamentally limited in such communities and simply asking for more labor from the moderators would likely be ineffective, if not inappropriate.} 

\mrev{Given this, the more constructive message we can derive from our findings would be that we need to better support the moderators so that whatever labor they are able to put in can yield the largest possible benefit to the community.} For instance, we could envision ways to provide better moderation tools that can more effectively flag content that needs to be reviewed by human moderators, both for its harmful content and for how exposed it might be (e.g., appearing at the top of a community’s feed). It also might be more tractable to train classifiers to identify the suspected type of violation in the moderators' AutoMod queues, so that moderators can more effectively triage. \mrev{Finally, it would also be critical to support the volunteer moderators in their work environment where they can be emotionally and psychologically strained~\cite{Matias2016, 35_Seering}, perhaps by constructing a protocol for common negative situations as suggested in prior work~\cite{Wohn2019}.}

\subsection{\mrev{Local Norms May Conflict With Macro Norms}}
\mrev{One caveat that is worth highlighting is how content from individual subreddits may not necessarily subscribe to all of the macro norms. That is, a macro norm violating comment might be acceptable in certain communities. For example, NSFW subreddits might be more lenient towards pornographic materials while political subreddits might be more lenient towards certain types of inflammatory political comments, which explains why topics on politics and NSFW are significant predictors of higher rates of violating comments. This is a tradeoff one has to make to measure the amount of violating content with a wide-lens across an entire platform. And although we believe that the eight categories of macro norms we used in our analysis give us a more nuanced understanding of what is going on in these communities than catch-all metrics (e.g., toxicity), we note that any work that aims to take such an approach needs to be cautious so as to not unfairly flag minority communities as being more violating~ \cite{Oliva2020} just because they deviate from certain macro norms.}

\mrev{In the scope of our work, we do find that even when excluding communities in topic categories such as NSFW and politics, the rate of violating comments remains high. This suggests that although some of the comments marked by our method as macro norm violating may be accepted by moderators within certain subreddits, this does not account for the majority of such comments. However, the estimations presented here should still be interpreted primarily as an attempt to understand the aggregate trend of norm violations across a large number of communities rather than to single out any one community as particularly problematic. This tension between the local norms and the macro norms should continue to be discussed in the future literature, importantly by directly communicating with the moderators and the members of their communities to better understand and reconcile the differences in between these two classes of norms.}

\subsection{Limitations and Future Work}
Our analysis is a wide-lens, macro-scale approach. However, such an analysis makes inevitable tradeoffs compared to deeper qualitative and inductive, or even more focused measurement, studies. Firstly, we focused on some of the most popular subreddits in 2016--2017. When studying the 2020 period, we stayed with these subreddits as we were interested in a longitudinal study of how these subreddits developed over time, but future work could investigate how 2020's most popular subreddits perform. It is possible that the reduced violation rate in 2020 could be attributed to subreddits losing popularity and trolls migrating elsewhere to the newer popular subreddits. We also restricted our analyses to the top 100 most popular subreddits, which is consistent with prior work but prevents us from measuring prevalence in smaller communities. An open question remains as to whether these smaller communities better regulate violating comments, or do less because they are not in the spotlight. 

In addition, our work contains methodological limitations that future work could expand upon. For instance, we focused on estimating the prevalence of violating comments. But one could also approach this problem by measuring the content view of such comments. That is, how often are violating comments seen by the users or are they mostly ignored and placed at the edge of the comment section? Here, we only indirectly touched on this topic by showing the significant difference in engagement between the moderated and unmoderated violating comments as engagement often dictates where a given comment is placed within the comment section. Also, we operationalized violations as what moderators would remove, but there may be more normative definitions that would produce different estimates. Future work should continue to rigorously define and operationalize violating comments. One important venue of such future work would work directly with the moderators and their communities to consider what the members of that particular community would consider to be a violation of their standards and choose to remove. Additionally, our classifiers could introduce bias or stale outputs in part because the training data is a few years old. We believe this is largely accounted for as the classifiers maintained high recall and their outputs were reviewed by human annotators. But developing better performing algorithms for identifying violating comments remains an important but also a challenging task. 

Finally, it is worth noting that a limitation to our estimation approach, which centers around human annotators, is the human cost. In many cases, employing annotators for a large number of hours may not be possible, particularly for an academic researcher, due to the limited resources available. However, we believe that it is important for non-industry academics to audit and inspect the state of social computing platforms. In our study, we overcome some of these practical challenges by combining human annotations with bootstrap resampling but we hope future work will continue to build on this precedence and refine techniques for large scale measurement studies.

%% file: content_minor_revision__Apr2022/conclusion.tex
\section{CONCLUSION}
In this work, we aimed to quantify and characterize norm violating comments on a major online community platform. To accomplish this, we defined norm violating comments as those that would have been removed on most of the communities, and developed a human-AI pipeline for identifying such comments at scale. Finally, we employed bootstrap estimation to estimate the prevalence of these comments from two distinct periods: one from 2016 May--2017 March, and another during the final three weeks of December 2020. In doing so, our work presents a model pipeline that could serve as a point of reference to future work that tries to identify violating comments and provides empirical results estimating the prevalence of violating comments. Our findings suggest that despite the growing efforts to reduce harmful content in online social spaces, a large number of violating comments still populate these communities. Based on our findings, we highlight the need for continuous effort to improve our designs, policies, and moderation strategies.

%% file: content_minor_revision__Apr2022/appendix.tex
\section{Details about the classifiers' training dataset}
\subsection{\rnr{Size of the Moderated Comments Dataset}}

\input{content_minor_revision__Apr2022/table/subreddit_size}

%% file: content_minor_revision__Apr2022/table/subreddit_size.tex
\noindent\begin{minipage}{\linewidth}
\centering
\begin{tabular}{ll;{1pt/1pt}ll;{1pt/1pt}ll} 
\hline\hline
Subreddits         & \multicolumn{1}{l}{Count} & Subreddits           & \multicolumn{1}{l}{Count} & Subreddits          & Count  \\ 
\toprule
2007scape          & 8335                      & gameofthrones        & 7136                      & pcmasterrace        & 16986  \\
Android            & 10650                     & Games                & 28587                     & personalfinance     & 22841  \\
anime              & 8365                      & gaming               & 12270                     & philosophy          & 8710   \\
AskHistorians      & 28772                     & GetMotivated         & 5823                      & photoshopbattles    & 19620  \\
AskReddit          & 110k                      & gifs                 & 10075                     & pics                & 16921  \\
askscience         & 38851                     & GlobalOffensive      & 16340                     & pokemon             & 8759   \\
AskTrumpSupporters & 7039                      & GlobalOffensiveTrade & 12751                     & pokemongo           & 16789  \\
AskWomen           & 13192                     & gonewild             & 51012                     & pokemontrades       & 14144  \\
asoiaf             & 5817                      & hearthstone          & 6335                      & PoliticalDiscussion & 26360  \\
atheism            & 10538                     & hillaryclinton       & 39683                     & politics            & 148k   \\
aww                & 21222                     & hiphopheads          & 9279                      & PurplePillDebate    & 5852   \\
BlackPeopleTwitter & 14200                     & history              & 13554                     & relationships       & 52987  \\
books              & 9501                      & IAmA                 & 6030                      & SandersForPresident & 16012  \\
canada             & 7742                      & india                & 9688                      & science             & 105k   \\
CanadaPolitics     & 7529                      & jailbreak            & 5616                      & sex                 & 12873  \\
CFB                & 17108                     & LateStageCapitalism  & 12303                     & ShitRedditSays      & 7798   \\
changemyview       & 8034                      & leagueoflegends      & 33035                     & Showerthoughts      & 11531  \\
Christianity       & 8578                      & legaladvice          & 13813                     & socialism           & 7720   \\
churning           & 8044                      & LifeProTips          & 6827                      & space               & 7976   \\
conspiracy         & 7920                      & me\_irl              & 6075                      & spacex              & 6148   \\
creepyPMs          & 6941                      & MMA                  & 15546                     & SubredditDrama      & 6811   \\
dataisbeautiful    & 5922                      & movies               & 9720                      & SuicideWatch        & 6914   \\
depression         & 9500                      & nba                  & 8629                      & syriancivilwar      & 19618  \\
DestinyTheGame     & 5982                      & NeutralPolitics      & 5679                      & technology          & 6768   \\
DIY                & 11187                     & news                 & 76660                     & television          & 5963   \\
EnoughTrumpSpam    & 14203                     & nfl                  & 14630                     & TheSilphRoad        & 6629   \\
europe             & 15261                     & nosleep              & 18335                     & tifu                & 8084   \\
explainlikeimfive  & 56100                     & nottheonion          & 7442                      & TwoXChromosomes     & 51083  \\
fantasyfootball    & 8149                      & NSFW\_GIF            & 5921                      & UpliftingNews       & 10521  \\
food               & 9960                      & OldSchoolCool        & 7048                      & videos              & 13510  \\
funny              & 16344                     & OutOfTheLoop         & 10573                     & whatisthisthing     & 11691  \\
Futurology         & 16213                     & Overwatch            & 8131                      & worldnews           & 95026  \\
wow                & \multicolumn{1}{l}{6896}  &                      & \multicolumn{1}{l}{}      &                     &        \\
\midrule

\end{tabular}
\captionof{table}{\rnr{The size of the moderated comment datasets per subreddit that were used to train the subreddit classifiers. In addition to the moderated comments, we gathered the same number of unmoderated comments for each of the subreddits to create a balanced dataset for training the classifiers.}}

\end{minipage}

%% file: main.bbl

\begin{thebibliography}{99}


\ifx \showCODEN    \undefined \def \showCODEN     #1{\unskip}     \fi
\ifx \showDOI      \undefined \def \showDOI       #1{#1}\fi
\ifx \showISBNx    \undefined \def \showISBNx     #1{\unskip}     \fi
\ifx \showISBNxiii \undefined \def \showISBNxiii  #1{\unskip}     \fi
\ifx \showISSN     \undefined \def \showISSN      #1{\unskip}     \fi
\ifx \showLCCN     \undefined \def \showLCCN      #1{\unskip}     \fi
\ifx \shownote     \undefined \def \shownote      #1{#1}          \fi
\ifx \showarticletitle \undefined \def \showarticletitle #1{#1}   \fi
\ifx \showURL      \undefined \def \showURL       {\relax}        \fi
\providecommand\bibfield[2]{#2}
\providecommand\bibinfo[2]{#2}
\providecommand\natexlab[1]{#1}
\providecommand\showeprint[2][]{arXiv:#2}

\bibitem[\protect\citeauthoryear{Alkhatib and Bernstein}{Alkhatib and
  Bernstein}{2019}]%
        {alkhatib2019street}
\bibfield{author}{\bibinfo{person}{Ali Alkhatib} {and} \bibinfo{person}{Michael
  Bernstein}.} \bibinfo{year}{2019}\natexlab{}.
\newblock \showarticletitle{Street-level algorithms: A theory at the gaps
  between policy and decisions}. In \bibinfo{booktitle}{\emph{Proceedings of
  the 2019 CHI Conference on Human Factors in Computing Systems}}.
  \bibinfo{pages}{1--13}.
\newblock


\bibitem[\protect\citeauthoryear{Anderson}{Anderson}{2019}]%
        {94_Anderson}
\bibfield{author}{\bibinfo{person}{Carolyn~J. Anderson}.}
  \bibinfo{year}{2019}\natexlab{}.
\newblock \bibinfo{booktitle}{\emph{Poisson Regression for Regression of Counts
  and Rates}}.
\newblock
\urldef\tempurl%
\url{https://education.illinois.edu/docs/default-source/carolyn-anderson/edpsy589/lectures/4_glm/4glm_3_beamer_post.pdf}
\showURL{%
Retrieved July 1, 2021 from \tempurl}


\bibitem[\protect\citeauthoryear{Angeli, Tibshirani, Wu, and Manning}{Angeli
  et~al\mbox{.}}{2014}]%
        {22_Angeli}
\bibfield{author}{\bibinfo{person}{Gabor Angeli}, \bibinfo{person}{Julie
  Tibshirani}, \bibinfo{person}{Jean Wu}, {and} \bibinfo{person}{Christopher~D.
  Manning}.} \bibinfo{year}{2014}\natexlab{}.
\newblock \showarticletitle{Combining Distant and Partial Supervision for
  Relation Extraction}. In \bibinfo{booktitle}{\emph{Proceedings of the 2014
  Conference on Empirical Methods in Natural Language Processing (EMNLP)}}.
  \bibinfo{publisher}{Association for Computational Linguistics},
  \bibinfo{pages}{1556--1567}.
\newblock
\urldef\tempurl%
\url{https://www.aclweb.org/anthology/D14-1164/}
\showURL{%
\tempurl}


\bibitem[\protect\citeauthoryear{Barocas, Crawford, Shapiro, and
  Wallach}{Barocas et~al\mbox{.}}{2017}]%
        {57_Barocas}
\bibfield{author}{\bibinfo{person}{Solon Barocas}, \bibinfo{person}{Kate
  Crawford}, \bibinfo{person}{Aaron Shapiro}, {and} \bibinfo{person}{Hanna
  Wallach}.} \bibinfo{year}{2017}\natexlab{}.
\newblock \showarticletitle{The Problem With Bias: Allocative Versus
  Representational Harms in Machine Learning}. In
  \bibinfo{booktitle}{\emph{SIGCIS Conference,
  http://meetings.sigcis.org/uploads/6/3/6/8/6368912/program.pdf}}.
  \bibinfo{publisher}{ACM}.
\newblock


\bibitem[\protect\citeauthoryear{Barrett}{Barrett}{2020}]%
        {91_Barrett}
\bibfield{author}{\bibinfo{person}{Paul~M. Barrett}.}
  \bibinfo{year}{2020}\natexlab{}.
\newblock \bibinfo{booktitle}{\emph{Who Moderates the Social Media Giants? A
  Call to End Outsourcing}}.
\newblock
\urldef\tempurl%
\url{https://bhr.stern.nyu.edu/tech-content-moderation-june-2020}
\showURL{%
Retrieved July 1, 2021 from \tempurl}


\bibitem[\protect\citeauthoryear{Bernstein, Monroy-Hernandez, Harry, Andre,
  Panovich, and Vargas}{Bernstein et~al\mbox{.}}{2011}]%
        {30_MSB}
\bibfield{author}{\bibinfo{person}{Michael~S Bernstein},
  \bibinfo{person}{Andres Monroy-Hernandez}, \bibinfo{person}{Drew Harry},
  \bibinfo{person}{Paul Andre}, \bibinfo{person}{Katrina Panovich}, {and}
  \bibinfo{person}{Gregory~G Vargas}.} \bibinfo{year}{2011}\natexlab{}.
\newblock \showarticletitle{4chan and/b: An Analysis of Anonymity and
  Ephemerality in a Large Online Community}. In \bibinfo{booktitle}{\emph{In
  ICWSM}}. \bibinfo{publisher}{AAAI}.
\newblock


\bibitem[\protect\citeauthoryear{Bickert}{Bickert}{2018}]%
        {39_Bickert}
\bibfield{author}{\bibinfo{person}{Monika Bickert}.}
  \bibinfo{year}{2018}\natexlab{}.
\newblock \bibinfo{booktitle}{\emph{Publishing Our Internal Enforcement
  Guidelines and Expanding Our Appeals Process}}.
\newblock
\urldef\tempurl%
\url{https://newsroom.fb.com/news/2018/04/comprehensive-community-standards/}
\showURL{%
Retrieved Dec 1, 2020 from \tempurl}


\bibitem[\protect\citeauthoryear{Binns, Veale, Kleek, and Shadbolt}{Binns
  et~al\mbox{.}}{2017}]%
        {10_Binns}
\bibfield{author}{\bibinfo{person}{Reuben Binns}, \bibinfo{person}{Michael
  Veale}, \bibinfo{person}{Max~Van Kleek}, {and} \bibinfo{person}{Nigel
  Shadbolt}.} \bibinfo{year}{2017}\natexlab{}.
\newblock \showarticletitle{Like Trainer, Like Bot? Inheritance of Bias in
  Algorithmic Content Moderation}. In \bibinfo{booktitle}{\emph{International
  Conference on Social Informatics}}. \bibinfo{publisher}{Springer},
  \bibinfo{address}{Springer, New York}, \bibinfo{pages}{405--415}.
\newblock
\urldef\tempurl%
\url{https://link.springer.com/chapter/10.1007/978-3-319-67256-4_32}
\showURL{%
\tempurl}


\bibitem[\protect\citeauthoryear{Blodgett, Green, and O'Connor}{Blodgett
  et~al\mbox{.}}{2016}]%
        {55_Blodgett}
\bibfield{author}{\bibinfo{person}{Su~Lin Blodgett}, \bibinfo{person}{Lisa
  Green}, {and} \bibinfo{person}{Brendan O'Connor}.}
  \bibinfo{year}{2016}\natexlab{}.
\newblock \showarticletitle{Demographic dialectal variation in social media: A
  case study of African-American English}. In \bibinfo{booktitle}{\emph{In:
  EMNLP 2016: Conference on Empirical Methods in Natural Language Processing}}.
  \bibinfo{publisher}{ACM}.
\newblock


\bibitem[\protect\citeauthoryear{Bor and Petersen}{Bor and Petersen}{2021}]%
        {bor_hostility}
\bibfield{author}{\bibinfo{person}{Alexander Bor} {and}
  \bibinfo{person}{Michael Petersen}.} \bibinfo{year}{2021}\natexlab{}.
\newblock \showarticletitle{The Psychology of Online Political Hostility: A
  Comprehensive, Cross-National Test of the Mismatch Hypothesis}.
\newblock \bibinfo{journal}{\emph{American Political Science Review}}
  \bibinfo{number}{APSR} (\bibinfo{year}{2021}).
\newblock


\bibitem[\protect\citeauthoryear{BOR and PETERSEN}{BOR and PETERSEN}{2021}]%
        {bor_petersen_2021}
\bibfield{author}{\bibinfo{person}{ALEXANDER BOR} {and}
  \bibinfo{person}{MICHAEL~BANG PETERSEN}.} \bibinfo{year}{2021}\natexlab{}.
\newblock \showarticletitle{The Psychology of Online Political Hostility: A
  Comprehensive, Cross-National Test of the Mismatch Hypothesis}.
\newblock \bibinfo{journal}{\emph{American Political Science Review}}
  (\bibinfo{year}{2021}), \bibinfo{pages}{1–18}.
\newblock
\urldef\tempurl%
\url{https://doi.org/10.1017/S0003055421000885}
\showDOI{\tempurl}


\bibitem[\protect\citeauthoryear{Brody and Diakopoulos}{Brody and
  Diakopoulos}{2011}]%
        {50_Brody}
\bibfield{author}{\bibinfo{person}{Samuel Brody} {and}
  \bibinfo{person}{Nicholas Diakopoulos}.} \bibinfo{year}{2011}\natexlab{}.
\newblock \showarticletitle{Cooooooooooooooollllllllllllll!!!!!!!!!!!!!!: using
  word lengthening to detect sentiment in microblogs}. In
  \bibinfo{booktitle}{\emph{In Proceedings of the conference on empirical
  methods in natural language processing}}. \bibinfo{publisher}{Association for
  Computational Linguistics}.
\newblock


\bibitem[\protect\citeauthoryear{Buckels, Trapnell, and Paulhus}{Buckels
  et~al\mbox{.}}{2014}]%
        {buckels2014trolls}
\bibfield{author}{\bibinfo{person}{Erin~E Buckels}, \bibinfo{person}{Paul~D
  Trapnell}, {and} \bibinfo{person}{Delroy~L Paulhus}.}
  \bibinfo{year}{2014}\natexlab{}.
\newblock \showarticletitle{Trolls just want to have fun}.
\newblock \bibinfo{journal}{\emph{Personality and individual Differences}}
  \bibinfo{volume}{67} (\bibinfo{year}{2014}), \bibinfo{pages}{97--102}.
\newblock


\bibitem[\protect\citeauthoryear{Butler, Joyce, and Pike}{Butler
  et~al\mbox{.}}{2008}]%
        {14_Butler}
\bibfield{author}{\bibinfo{person}{Brian Butler}, \bibinfo{person}{Elisabeth
  Joyce}, {and} \bibinfo{person}{Jacqueline Pike}.}
  \bibinfo{year}{2008}\natexlab{}.
\newblock \showarticletitle{Don't look now, but we've created a bureaucracy:
  the nature and roles of policies and rules in wikipedia}. In
  \bibinfo{booktitle}{\emph{CHI '08: Proceedings of the SIGCHI Conference on
  Human Factors in Computing Systems}}. \bibinfo{publisher}{ACM},
  \bibinfo{pages}{1101–1110}.
\newblock
\urldef\tempurl%
\url{https://doi.org/10.1145/1357054.1357227}
\showDOI{\tempurl}


\bibitem[\protect\citeauthoryear{by~Jim}{by~Jim}{2020}]%
        {81_Bootstrap}
\bibfield{author}{\bibinfo{person}{Statistics by Jim}.}
  \bibinfo{year}{2020}\natexlab{}.
\newblock \bibinfo{booktitle}{\emph{Introduction to Bootstrapping in Statistics
  with an Example}}.
\newblock
\urldef\tempurl%
\url{https://statisticsbyjim.com/hypothesis-testing/bootstrapping/}
\showURL{%
Retrieved Dec 1, 2020 from \tempurl}


\bibitem[\protect\citeauthoryear{Chancellor, Kalantidis, Pater, Choudhury, and
  Shamma}{Chancellor et~al\mbox{.}}{2017}]%
        {93_Chancellor}
\bibfield{author}{\bibinfo{person}{Stevie Chancellor}, \bibinfo{person}{Yannis
  Kalantidis}, \bibinfo{person}{Jessica Pater}, \bibinfo{person}{Munmun~De
  Choudhury}, {and} \bibinfo{person}{David~A. Shamma}.}
  \bibinfo{year}{2017}\natexlab{}.
\newblock \showarticletitle{Multimodal Classification of Moderated Online
  Pro-Eating Disorder Content}.
\newblock \bibinfo{journal}{\emph{Proceedings of the 2017 CHI Conference on
  Human Factors in Computing Systems}} (\bibinfo{year}{2017}).
\newblock


\bibitem[\protect\citeauthoryear{Chancellor, Lin, and Choudhury}{Chancellor
  et~al\mbox{.}}{2016a}]%
        {4_Chancellor}
\bibfield{author}{\bibinfo{person}{Stevie Chancellor},
  \bibinfo{person}{Zhiyuan~Jerry Lin}, {and} \bibinfo{person}{Munmun~De
  Choudhury}.} \bibinfo{year}{2016}\natexlab{a}.
\newblock \showarticletitle{``This Post Will Just Get Taken Down'':
  Characterizing Removed Pro-Eating Disorder Social Media Content}. In
  \bibinfo{booktitle}{\emph{Proceedings of the 2016 CHI Conference on Human
  Factors in Computing Systems}}. \bibinfo{publisher}{ACM},
  \bibinfo{address}{San Jose, CA, USA}, \bibinfo{pages}{1157–--1162}.
\newblock
\urldef\tempurl%
\url{https://doi.org/10.1145/2858036.2858248}
\showDOI{\tempurl}


\bibitem[\protect\citeauthoryear{Chancellor, Pater, Clear, Gilbert, and
  Choudhury}{Chancellor et~al\mbox{.}}{2016b}]%
        {51_Chancellor}
\bibfield{author}{\bibinfo{person}{Stevie Chancellor},
  \bibinfo{person}{Jessica~Annette Pater}, \bibinfo{person}{Trustin Clear},
  \bibinfo{person}{Eric Gilbert}, {and} \bibinfo{person}{Munmun~De Choudhury}.}
  \bibinfo{year}{2016}\natexlab{b}.
\newblock \showarticletitle{\# thyghgapp: Instagram Content Moderation and
  Lexical Variation in Pro-Eating Disorder Communities}. In
  \bibinfo{booktitle}{\emph{In Proceedings of the 19th ACM Conference on
  Computer-Supported Cooperative Work \& Social Computing}}.
  \bibinfo{publisher}{ACM}.
\newblock


\bibitem[\protect\citeauthoryear{Chandrasekharan, Gandhi, Mustelier, and
  Gilbert}{Chandrasekharan et~al\mbox{.}}{2019}]%
        {chandrasekharan2019crossmod}
\bibfield{author}{\bibinfo{person}{Eshwar Chandrasekharan},
  \bibinfo{person}{Chaitrali Gandhi}, \bibinfo{person}{Matthew~Wortley
  Mustelier}, {and} \bibinfo{person}{Eric Gilbert}.}
  \bibinfo{year}{2019}\natexlab{}.
\newblock \showarticletitle{Crossmod: A Cross-Community Learning-Based System
  to Assist Reddit Moderators}.
\newblock \bibinfo{journal}{\emph{Proc. ACM Hum.-Comput. Interact.}}
  \bibinfo{volume}{3}, \bibinfo{number}{CSCW}, Article \bibinfo{articleno}{174}
  (\bibinfo{date}{Nov.} \bibinfo{year}{2019}), \bibinfo{numpages}{30}~pages.
\newblock
\urldef\tempurl%
\url{https://doi.org/10.1145/3359276}
\showDOI{\tempurl}


\bibitem[\protect\citeauthoryear{Chandrasekharan and Gilbert}{Chandrasekharan
  and Gilbert}{2019}]%
        {92_Chandrasekharan}
\bibfield{author}{\bibinfo{person}{Eshwar Chandrasekharan} {and}
  \bibinfo{person}{Eric Gilbert}.} \bibinfo{year}{2019}\natexlab{}.
\newblock \bibinfo{booktitle}{\emph{Hybrid Approaches to Detect Comments
  Violating Macro Norms on Reddit}}.
\newblock
\urldef\tempurl%
\url{https://arxiv.org/pdf/1904.03596.pdf}
\showURL{%
Retrieved July 1, 2021 from \tempurl}


\bibitem[\protect\citeauthoryear{Chandrasekharan, Pavalanathan, Srinivasan,
  Glynn, Eisenstein, and Gilbert}{Chandrasekharan et~al\mbox{.}}{2017a}]%
        {7_Chandrasekharan}
\bibfield{author}{\bibinfo{person}{Eshwar Chandrasekharan},
  \bibinfo{person}{Umashanthi Pavalanathan}, \bibinfo{person}{Anirudh
  Srinivasan}, \bibinfo{person}{Adam Glynn}, \bibinfo{person}{Jacob
  Eisenstein}, {and} \bibinfo{person}{Eric Gilbert}.}
  \bibinfo{year}{2017}\natexlab{a}.
\newblock \showarticletitle{You Can't Stay Here: The Efficacy of Reddit's 2015
  Ban Examined Through Hate Speech}.
\newblock \bibinfo{journal}{\emph{Proc. ACM Hum.-Comput. Interact.}}
  (\bibinfo{year}{2017}), \bibinfo{pages}{22 pages}.
\newblock
\urldef\tempurl%
\url{https://dl.acm.org/doi/10.1145/3134666}
\showURL{%
\tempurl}


\bibitem[\protect\citeauthoryear{Chandrasekharan, Samory, Jhaver, Charvat,
  Bruckman, Lampe, Eisenstein, and Gilbert}{Chandrasekharan
  et~al\mbox{.}}{2018}]%
        {Chandrasekharan2018internet}
\bibfield{author}{\bibinfo{person}{Eshwar Chandrasekharan},
  \bibinfo{person}{Mattia Samory}, \bibinfo{person}{Shagun Jhaver},
  \bibinfo{person}{Hunter Charvat}, \bibinfo{person}{Amy Bruckman},
  \bibinfo{person}{Cliff Lampe}, \bibinfo{person}{Jacob Eisenstein}, {and}
  \bibinfo{person}{Eric Gilbert}.} \bibinfo{year}{2018}\natexlab{}.
\newblock \showarticletitle{The Internet's Hidden Rules: An Empirical Study of
  Reddit Norm Violations at Micro, Meso, and Macro Scales}.
\newblock \bibinfo{journal}{\emph{Proc. ACM Hum.-Comput. Interact.}}
  \bibinfo{volume}{2}, \bibinfo{number}{CSCW}, Article \bibinfo{articleno}{32}
  (\bibinfo{date}{Nov.} \bibinfo{year}{2018}), \bibinfo{numpages}{25}~pages.
\newblock
\showISSN{2573-0142}
\urldef\tempurl%
\url{https://doi.org/10.1145/3274301}
\showDOI{\tempurl}


\bibitem[\protect\citeauthoryear{Chandrasekharan, Samory, Srinivasan, and
  Gilbert}{Chandrasekharan et~al\mbox{.}}{2017b}]%
        {1_Chandrasekharan}
\bibfield{author}{\bibinfo{person}{Eshwar Chandrasekharan},
  \bibinfo{person}{Mattia Samory}, \bibinfo{person}{Anirudh Srinivasan}, {and}
  \bibinfo{person}{Eric Gilbert}.} \bibinfo{year}{2017}\natexlab{b}.
\newblock \showarticletitle{The Bag of Communities: Identifying Abusive
  Behavior Online with Preexisting Internet Data}. In
  \bibinfo{booktitle}{\emph{Proceedings of the 2017 CHI Conference on Human
  Factors in Computing Systems}}. \bibinfo{publisher}{ACM},
  \bibinfo{address}{Denver, CO, USA}, \bibinfo{pages}{3175--3187}.
\newblock
\urldef\tempurl%
\url{https://doi.org/10.1145/3025453.3026018}
\showDOI{\tempurl}


\bibitem[\protect\citeauthoryear{Chang, Cheng, and
  Danescu-Niculescu-Mizil}{Chang et~al\mbox{.}}{2020}]%
        {chang2020don}
\bibfield{author}{\bibinfo{person}{Jonathan~P Chang}, \bibinfo{person}{Justin
  Cheng}, {and} \bibinfo{person}{Cristian Danescu-Niculescu-Mizil}.}
  \bibinfo{year}{2020}\natexlab{}.
\newblock \showarticletitle{Don’t let me be misunderstood: Comparing
  intentions and perceptions in online discussions}. In
  \bibinfo{booktitle}{\emph{Proceedings of The Web Conference 2020}}.
  \bibinfo{pages}{2066--2077}.
\newblock


\bibitem[\protect\citeauthoryear{Cheng, Bernstein, Danescu-Niculescu-Mizil, and
  Leskovec}{Cheng et~al\mbox{.}}{2017}]%
        {cheng2017anyone}
\bibfield{author}{\bibinfo{person}{Justin Cheng}, \bibinfo{person}{Michael
  Bernstein}, \bibinfo{person}{Cristian Danescu-Niculescu-Mizil}, {and}
  \bibinfo{person}{Jure Leskovec}.} \bibinfo{year}{2017}\natexlab{}.
\newblock \showarticletitle{Anyone Can Become a Troll: Causes of Trolling
  Behavior in Online Discussions}. In \bibinfo{booktitle}{\emph{Proceedings of
  the 2017 ACM Conference on Computer Supported Cooperative Work and Social
  Computing}} (Portland, Oregon, USA) \emph{(\bibinfo{series}{CSCW '17})}.
  \bibinfo{publisher}{ACM}, \bibinfo{address}{New York, NY, USA},
  \bibinfo{pages}{1217--1230}.
\newblock
\showISBNx{978-1-4503-4335-0}
\urldef\tempurl%
\url{https://doi.org/10.1145/2998181.2998213}
\showDOI{\tempurl}


\bibitem[\protect\citeauthoryear{Cheng, Danescu-Niculescu-Mizil, and
  Leskovec}{Cheng et~al\mbox{.}}{2016}]%
        {5_Cheng}
\bibfield{author}{\bibinfo{person}{Justin Cheng}, \bibinfo{person}{Cristian
  Danescu-Niculescu-Mizil}, {and} \bibinfo{person}{Jure Leskovec}.}
  \bibinfo{year}{2016}\natexlab{}.
\newblock \showarticletitle{Antisocial Behavior in Online Discussion
  Communities}. In \bibinfo{booktitle}{\emph{In Ninth International AAAI
  Conference on Web and Social Media}}. \bibinfo{publisher}{AAAI}.
\newblock
\urldef\tempurl%
\url{https://cs.stanford.edu/people/jure/pubs/trolls-icwsm15.pdf}
\showURL{%
\tempurl}


\bibitem[\protect\citeauthoryear{Claire}{Claire}{2010}]%
        {82_Claire}
\bibfield{author}{\bibinfo{person}{Hardaker Claire}.}
  \bibinfo{year}{2010}\natexlab{}.
\newblock \showarticletitle{Trolling in asynchronous computer-mediated
  communication: From user discussions to academic definitions}.
\newblock \bibinfo{journal}{\emph{Politeness Res}} (\bibinfo{year}{2010}).
\newblock


\bibitem[\protect\citeauthoryear{Cranz and Brandom}{Cranz and Brandom}{2021}]%
        {facebook_scandal_2}
\bibfield{author}{\bibinfo{person}{Alex Cranz} {and} \bibinfo{person}{Russell
  Brandom}.} \bibinfo{year}{2021}\natexlab{}.
\newblock \showarticletitle{Facebook encourages hate speech for profit, says
  whistleblower}.
\newblock \bibinfo{journal}{\emph{The Verge}} (\bibinfo{year}{2021}).
\newblock


\bibitem[\protect\citeauthoryear{Culliford and Paul}{Culliford and
  Paul}{2020}]%
        {58_Culliford}
\bibfield{author}{\bibinfo{person}{Elizabeth Culliford} {and}
  \bibinfo{person}{Katie Paul}.} \bibinfo{year}{2020}\natexlab{}.
\newblock \bibinfo{booktitle}{\emph{Facebook offers up first-ever estimate of
  hate speech prevalence on its platform}}.
\newblock
\urldef\tempurl%
\url{https://www.reuters.com/article/us-facebook-content/facebook-estimates-hate-speech-seen-in-1-out-of-1000-views-on-its-platform-idUSKBN27Z2R0}
\showURL{%
Retrieved Dec 1, 2020 from \tempurl}


\bibitem[\protect\citeauthoryear{Diaz, Johnson, Lazar, Piper, and Gergle}{Diaz
  et~al\mbox{.}}{2018}]%
        {54_Diaz}
\bibfield{author}{\bibinfo{person}{Mark Diaz}, \bibinfo{person}{Isaac Johnson},
  \bibinfo{person}{Amanda Lazar}, \bibinfo{person}{Anne~Marie Piper}, {and}
  \bibinfo{person}{Darren Gergle}.} \bibinfo{year}{2018}\natexlab{}.
\newblock \showarticletitle{Addressing Age-Related Bias in Sentiment Analysis}.
  In \bibinfo{booktitle}{\emph{Proceedings of the 2018 CHI Conference on Human
  Factors in Computing Systems}}. \bibinfo{publisher}{ACM}.
\newblock


\bibitem[\protect\citeauthoryear{Dibbell}{Dibbell}{1993}]%
        {Dibbell1993rape}
\bibfield{author}{\bibinfo{person}{Julian Dibbell}.}
  \bibinfo{year}{1993}\natexlab{}.
\newblock \showarticletitle{A Rape in Cyberspace: How an Evil Clown, a Haitian
  Trickster Spirit, Two Wizards, and a Cast of Dozens Turned a Database Into a
  Society}.
\newblock \bibinfo{journal}{\emph{The Village Voice}}
  \bibinfo{volume}{December 23} (\bibinfo{year}{1993}),
  \bibinfo{pages}{36--42}.
\newblock
\urldef\tempurl%
\url{https://www.villagevoice.com/2005/10/18/a-rape-in-cyberspace/}
\showURL{%
\tempurl}


\bibitem[\protect\citeauthoryear{Dinakar, Reichart, and Lieberman}{Dinakar
  et~al\mbox{.}}{2011}]%
        {43_Dinakar}
\bibfield{author}{\bibinfo{person}{Karthik Dinakar}, \bibinfo{person}{Roi
  Reichart}, {and} \bibinfo{person}{Henry Lieberman}.}
  \bibinfo{year}{2011}\natexlab{}.
\newblock \showarticletitle{Modeling the detection of Textual Cyberbullying}.
  In \bibinfo{booktitle}{\emph{In The Social Mobile Web}}.
\newblock


\bibitem[\protect\citeauthoryear{Donovan}{Donovan}{2019}]%
        {2_Donovan}
\bibfield{author}{\bibinfo{person}{Joan Donovan}.}
  \bibinfo{year}{2019}\natexlab{}.
\newblock \showarticletitle{How Hate Groups' Secret Sound System Works}.
\newblock \bibinfo{journal}{\emph{The Atlantic}} (\bibinfo{date}{March}
  \bibinfo{year}{2019}).
\newblock
\urldef\tempurl%
\url{https://doi.org/10.1145/1188913.1188915}
\showDOI{\tempurl}


\bibitem[\protect\citeauthoryear{Fan and Zhang}{Fan and Zhang}{2020}]%
        {13_Fan}
\bibfield{author}{\bibinfo{person}{Jenny Fan} {and} \bibinfo{person}{Amy~X.
  Zhang}.} \bibinfo{year}{2020}\natexlab{}.
\newblock \showarticletitle{Digital Juries: A Civics-Oriented Approach to
  Platform Governance}. In \bibinfo{booktitle}{\emph{Conference on Human
  Factors in Computing Systems (CHI)}}. \bibinfo{publisher}{ACM},
  \bibinfo{pages}{1--14}.
\newblock
\urldef\tempurl%
\url{https://doi.org/10.1145/3313831.3376293}
\showDOI{\tempurl}


\bibitem[\protect\citeauthoryear{Gardner, Mulvey, and Shaw}{Gardner
  et~al\mbox{.}}{1995}]%
        {95_Gardner}
\bibfield{author}{\bibinfo{person}{Williamm Gardner}, \bibinfo{person}{Edward
  Mulvey}, {and} \bibinfo{person}{Esther Shaw}.}
  \bibinfo{year}{1995}\natexlab{}.
\newblock \showarticletitle{Regression Analyses of Counts and Rates: Poisson,
  Overdispersed Poisson, and Negative Binomial Models}.
\newblock \bibinfo{journal}{\emph{Quantitative Methods in Psychology}}
  (\bibinfo{year}{1995}).
\newblock


\bibitem[\protect\citeauthoryear{Geiger}{Geiger}{2011}]%
        {45_Geiger}
\bibfield{author}{\bibinfo{person}{R.Stuart Geiger}.}
  \bibinfo{year}{2011}\natexlab{}.
\newblock \bibinfo{booktitle}{\emph{The lives of bots. In: Wikipedia: A
  Critical Reader}}.
\newblock \bibinfo{publisher}{Amsterdam: Institute of Network CulturesSage}.
\newblock


\bibitem[\protect\citeauthoryear{GIFCT}{GIFCT}{2019}]%
        {46_GIFCT}
\bibfield{author}{\bibinfo{person}{GIFCT}.} \bibinfo{year}{2019}\natexlab{}.
\newblock \bibinfo{booktitle}{\emph{About the global internet forum to counter
  terrorism}}.
\newblock
\urldef\tempurl%
\url{https://perma.cc/44V5-554U}
\showURL{%
Retrieved Dec 1, 2020 from \tempurl}


\bibitem[\protect\citeauthoryear{Gilbert}{Gilbert}{2013}]%
        {18_Gilbert}
\bibfield{author}{\bibinfo{person}{Eric Gilbert}.}
  \bibinfo{year}{2013}\natexlab{}.
\newblock \showarticletitle{Widespread underprovision on Reddit}. In
  \bibinfo{booktitle}{\emph{Proceedings of the 2013 conference on Computer
  supported cooperative work (CSCW)}}. \bibinfo{publisher}{ACM},
  \bibinfo{pages}{803--808}.
\newblock
\urldef\tempurl%
\url{https://doi.org/10.1145/2441776.2441866}
\showDOI{\tempurl}


\bibitem[\protect\citeauthoryear{Gilbert}{Gilbert}{2020}]%
        {Gilbert2020}
\bibfield{author}{\bibinfo{person}{Sarah Gilbert}.}
  \bibinfo{year}{2020}\natexlab{}.
\newblock \showarticletitle{“I run the world’s largest historical outreach
  project and it’s on a cesspool of a website.” Moderating a public
  scholarship site on Reddit: A case study of r/AskHistorians}.
\newblock \bibinfo{journal}{\emph{Proceedings of the ACM on Human-Computer
  Interaction}} (\bibinfo{year}{2020}).
\newblock


\bibitem[\protect\citeauthoryear{Gillespie}{Gillespie}{2017}]%
        {34_Gillespie}
\bibfield{author}{\bibinfo{person}{Tarleton Gillespie}.}
  \bibinfo{year}{2017}\natexlab{}.
\newblock \bibinfo{booktitle}{\emph{Governance of and by platforms: Sage
  handbook of social media}}.
\newblock \bibinfo{publisher}{London: Sage}.
\newblock


\bibitem[\protect\citeauthoryear{Gillespie}{Gillespie}{2018}]%
        {8_Gillespie}
\bibfield{author}{\bibinfo{person}{Tarleton Gillespie}.}
  \bibinfo{year}{2018}\natexlab{}.
\newblock \bibinfo{booktitle}{\emph{Custodians of the Internet: Platforms,
  Content Moderation, and the Hidden Decisions that Shape Social Media}}.
\newblock \bibinfo{publisher}{Yale University Press}, \bibinfo{address}{New
  Haven, CT}.
\newblock


\bibitem[\protect\citeauthoryear{Google}{Google}{2018}]%
        {40_Google}
\bibfield{author}{\bibinfo{person}{Google}.} \bibinfo{year}{2018}\natexlab{}.
\newblock \bibinfo{booktitle}{\emph{YouTube Community Guidelines enforcement in
  Google's Tranparency Report for 2018}}.
\newblock
\urldef\tempurl%
\url{https://transparencyreport.google.com/youtube-policy/removals}
\showURL{%
Retrieved Dec 1, 2020 from \tempurl}


\bibitem[\protect\citeauthoryear{Gorwa, Binns, and Katzenbach}{Gorwa
  et~al\mbox{.}}{2020}]%
        {28_Gorwa}
\bibfield{author}{\bibinfo{person}{Robert Gorwa}, \bibinfo{person}{Reuben
  Binns}, {and} \bibinfo{person}{Christian Katzenbach}.}
  \bibinfo{year}{2020}\natexlab{}.
\newblock \showarticletitle{Algorithmic content moderation: Technical and
  political challenges in the automation of platform governance}.
\newblock \bibinfo{journal}{\emph{Big Data and Society}}
  (\bibinfo{year}{2020}).
\newblock
\urldef\tempurl%
\url{https://journals.sagepub.com/doi/full/10.1177/2053951719897945}
\showURL{%
\tempurl}


\bibitem[\protect\citeauthoryear{Hosseini, Kannan, Zhang, and
  Poovendran}{Hosseini et~al\mbox{.}}{2017}]%
        {11_Hosseini}
\bibfield{author}{\bibinfo{person}{Hossein Hosseini}, \bibinfo{person}{Sreeram
  Kannan}, \bibinfo{person}{Baosen Zhang}, {and} \bibinfo{person}{Radha
  Poovendran}.} \bibinfo{year}{2017}\natexlab{}.
\newblock \showarticletitle{Deceiving Google’s Perspective API Built for
  Detecting Toxic Comments}. In \bibinfo{booktitle}{\emph{arxiv}}.
\newblock
\urldef\tempurl%
\url{https://arxiv.org/pdf/1702.08138.pdf}
\showURL{%
\tempurl}


\bibitem[\protect\citeauthoryear{Kayany}{Kayany}{1998}]%
        {83_Kayany}
\bibfield{author}{\bibinfo{person}{Joseph~M. Kayany}.}
  \bibinfo{year}{1998}\natexlab{}.
\newblock \showarticletitle{Contexts of uninhibited online behavior: Flaming in
  social newsgroups on Usenet}.
\newblock \bibinfo{journal}{\emph{J Am Soc Inf Sci}} (\bibinfo{year}{1998}).
\newblock


\bibitem[\protect\citeauthoryear{Kiene, Shores, Chandrasekharan, Jhaver, Jiang,
  Dym, Seering, Gilbert, Lo, Wohn, and Dosono}{Kiene et~al\mbox{.}}{2019}]%
        {Kiene2019}
\bibfield{author}{\bibinfo{person}{Charles Kiene}, \bibinfo{person}{Kenny
  Shores}, \bibinfo{person}{Eshwar Chandrasekharan}, \bibinfo{person}{Shagun
  Jhaver}, \bibinfo{person}{Jialun Jiang}, \bibinfo{person}{Brianna Dym},
  \bibinfo{person}{Joseph Seering}, \bibinfo{person}{Sarah Gilbert},
  \bibinfo{person}{Kat Lo}, \bibinfo{person}{Donghee~Yvette Wohn}, {and}
  \bibinfo{person}{Bryan Dosono}.} \bibinfo{year}{2019}\natexlab{}.
\newblock \showarticletitle{Volunteer work: Mapping the future of moderation
  research}.
\newblock \bibinfo{journal}{\emph{Conference Companion Publication of the 2019
  on Computer Supported Cooperative Work and Social Computing}}
  (\bibinfo{year}{2019}).
\newblock


\bibitem[\protect\citeauthoryear{Kiesler, Kraut, Resnick, and Kittur}{Kiesler
  et~al\mbox{.}}{2012}]%
        {kiesler2012regulating}
\bibfield{author}{\bibinfo{person}{Sara Kiesler}, \bibinfo{person}{Robert
  Kraut}, \bibinfo{person}{Paul Resnick}, {and} \bibinfo{person}{Aniket
  Kittur}.} \bibinfo{year}{2012}\natexlab{}.
\newblock \showarticletitle{{Regulating behavior in online communities}}.
\newblock In \bibinfo{booktitle}{\emph{Building Successful Online Communities:
  Evidence-Based Social Design}}, \bibfield{editor}{\bibinfo{person}{Robert
  Kraut} {and} \bibinfo{person}{Paul Resnick}} (Eds.). \bibinfo{publisher}{MIT
  Press}, \bibinfo{address}{Cambridge, MA, USA}, Chapter~4,
  \bibinfo{pages}{125--177}.
\newblock


\bibitem[\protect\citeauthoryear{Kiesler, Siegel, and McGuire}{Kiesler
  et~al\mbox{.}}{1984}]%
        {Kiesler1984social}
\bibfield{author}{\bibinfo{person}{Sara Kiesler}, \bibinfo{person}{Jane
  Siegel}, {and} \bibinfo{person}{Timothy~W. McGuire}.}
  \bibinfo{year}{1984}\natexlab{}.
\newblock \showarticletitle{{Social psychological aspects of computer-mediated
  communication}}.
\newblock \bibinfo{journal}{\emph{American Psychologist}} \bibinfo{volume}{39},
  \bibinfo{number}{10} (\bibinfo{year}{1984}), \bibinfo{pages}{1123--1134}.
\newblock
\urldef\tempurl%
\url{https://doi.org/10.1037/0003-066X.39.10.1123}
\showDOI{\tempurl}


\bibitem[\protect\citeauthoryear{Klonick}{Klonick}{2017}]%
        {79_Klonick}
\bibfield{author}{\bibinfo{person}{Kate Klonick}.}
  \bibinfo{year}{2017}\natexlab{}.
\newblock \showarticletitle{The New Governors: The People, Rules, and Processes
  Governing Online Speech}.
\newblock \bibinfo{journal}{\emph{Harvard Law Review}} (\bibinfo{year}{2017}).
\newblock


\bibitem[\protect\citeauthoryear{Kraut and Resnick}{Kraut and Resnick}{2012}]%
        {88_Kraut}
\bibfield{author}{\bibinfo{person}{Robert~E. Kraut} {and} \bibinfo{person}{Paul
  Resnick}.} \bibinfo{year}{2012}\natexlab{}.
\newblock \bibinfo{booktitle}{\emph{Building successful online communities:
  Evidence-based social design}}.
\newblock \bibinfo{publisher}{MIT Press}, \bibinfo{address}{Cambridge,
  Massachusetts, United States}.
\newblock


\bibitem[\protect\citeauthoryear{Lampe and Resnick}{Lampe and Resnick}{2004}]%
        {27_Lampe}
\bibfield{author}{\bibinfo{person}{Cliff Lampe} {and} \bibinfo{person}{Paul
  Resnick}.} \bibinfo{year}{2004}\natexlab{}.
\newblock \showarticletitle{Slash(dot) and Burn: Distributed Moderation in a
  Large Online Conversation Space}. In \bibinfo{booktitle}{\emph{Proc. of ACM
  Computer Human Interaction Conference 2004 (CHI)}}.
  \bibinfo{publisher}{Association for Computing Machinery}.
\newblock
\urldef\tempurl%
\url{http://www.presnick.people.si.umich.edu/papers/chi04/LampeResnick.pdf}
\showURL{%
\tempurl}


\bibitem[\protect\citeauthoryear{Laub}{Laub}{2019}]%
        {71_Laub}
\bibfield{author}{\bibinfo{person}{Zachary Laub}.}
  \bibinfo{year}{2019}\natexlab{}.
\newblock \bibinfo{booktitle}{\emph{Hate Speech on Social Media: Global
  Comparisons}}.
\newblock
\urldef\tempurl%
\url{https://www.cfr.org/backgrounder/hate-speech-social-media-global-comparisons}
\showURL{%
Retrieved Dec 1, 2020 from \tempurl}


\bibitem[\protect\citeauthoryear{Lessig}{Lessig}{1999}]%
        {31_Lessig}
\bibfield{author}{\bibinfo{person}{Lawrence Lessig}.}
  \bibinfo{year}{1999}\natexlab{}.
\newblock \bibinfo{booktitle}{\emph{Code and other laws of cyberspace}}.
\newblock \bibinfo{publisher}{Basic books New York}.
\newblock


\bibitem[\protect\citeauthoryear{Li and J}{Li and J}{2018}]%
        {52_li}
\bibfield{author}{\bibinfo{person}{S Li} {and} \bibinfo{person}{Williams J}.}
  \bibinfo{year}{2018}\natexlab{}.
\newblock \bibinfo{booktitle}{\emph{Despite What Zuckerberg's Testimony May
  Imply, AI Cannot Save Us. Electronic Frontier Foundation Deeplinks Blog}}.
\newblock
\urldef\tempurl%
\url{https://www.eff.org/deeplinks/2018/04/despite-what-zuckerbergs-testimony-may-imply-ai-cannot-save-us}
\showURL{%
Retrieved Dec 1, 2020 from \tempurl}


\bibitem[\protect\citeauthoryear{Liu, Soderland, Bragg, Lin, Ling, and
  Weld}{Liu et~al\mbox{.}}{2016}]%
        {25_Liu}
\bibfield{author}{\bibinfo{person}{Angli Liu}, \bibinfo{person}{Stephen
  Soderland}, \bibinfo{person}{Jonathan Bragg}, \bibinfo{person}{Christopher~H.
  Lin}, \bibinfo{person}{Xiao Ling}, {and} \bibinfo{person}{Daniel~S. Weld}.}
  \bibinfo{year}{2016}\natexlab{}.
\newblock \showarticletitle{Effective Crowd Annotation for Relation
  Extraction}. In \bibinfo{booktitle}{\emph{Proceedings of the 2016 Conference
  of the North American Chapter of the Association for Computational
  Linguistics: Human Language Technologies}}. \bibinfo{publisher}{Association
  for Computational Linguistics}, \bibinfo{pages}{897--906}.
\newblock
\urldef\tempurl%
\url{https://www.aclweb.org/anthology/N16-1104/}
\showURL{%
\tempurl}


\bibitem[\protect\citeauthoryear{Matias}{Matias}{2016}]%
        {Matias2016}
\bibfield{author}{\bibinfo{person}{J.~Nathan Matias}.}
  \bibinfo{year}{2016}\natexlab{}.
\newblock \showarticletitle{Going dark: Social factors in collective action
  against platform operators in the Reddit blackout}.
\newblock \bibinfo{journal}{\emph{Proceedings of the 2016 CHI Conference on
  Human Factors in Computing Systems}} (\bibinfo{year}{2016}).
\newblock


\bibitem[\protect\citeauthoryear{Oliva, Antonialli, and Gomes}{Oliva
  et~al\mbox{.}}{2020}]%
        {Oliva2020}
\bibfield{author}{\bibinfo{person}{Thiago~Dias Oliva},
  \bibinfo{person}{Dennys~Marcelo Antonialli}, {and}
  \bibinfo{person}{Alessandra Gomes}.} \bibinfo{year}{2020}\natexlab{}.
\newblock \showarticletitle{Fighting Hate Speech, Silencing Drag Queens?
  Artificial Intelligence in Content Moderation and Risks to LGBTQ Voices
  Online}.
\newblock \bibinfo{journal}{\emph{Sexuality \& Culture}}
  (\bibinfo{year}{2020}).
\newblock


\bibitem[\protect\citeauthoryear{Pater, Nadji, Mynatt, and Bruckman}{Pater
  et~al\mbox{.}}{2014}]%
        {48_Pater}
\bibfield{author}{\bibinfo{person}{Jessica~Annette Pater},
  \bibinfo{person}{Yacin Nadji}, \bibinfo{person}{Elizabeth~D Mynatt}, {and}
  \bibinfo{person}{Amy~S Bruckman}.} \bibinfo{year}{2014}\natexlab{}.
\newblock \showarticletitle{Just awful enough: the functional dysfunction of
  the something awful forums}. In \bibinfo{booktitle}{\emph{In Proceedings of
  the 32nd annual ACM conference on Human factors in computing systems, ACM}}.
  \bibinfo{publisher}{ACM}.
\newblock


\bibitem[\protect\citeauthoryear{Pelley}{Pelley}{2021}]%
        {facebook_scandal}
\bibfield{author}{\bibinfo{person}{Scott Pelley}.}
  \bibinfo{year}{2021}\natexlab{}.
\newblock \bibinfo{title}{Whistleblower: Facebook is misleading the public on
  progress against hate speech, violence, misinformation}.
\newblock
\newblock
\urldef\tempurl%
\url{https://www.cbsnews.com/news/facebook-whistleblower-frances-haugen-misinformation-public-60-minutes-2021-10-03/}
\showURL{%
\tempurl}


\bibitem[\protect\citeauthoryear{Pershina, Min, Xu, and Grishman}{Pershina
  et~al\mbox{.}}{2014}]%
        {23_Pershina}
\bibfield{author}{\bibinfo{person}{Maria Pershina}, \bibinfo{person}{Bonan
  Min}, \bibinfo{person}{Wei Xu}, {and} \bibinfo{person}{Ralph Grishman}.}
  \bibinfo{year}{2014}\natexlab{}.
\newblock \showarticletitle{Infusion of Labeled Data into Distant Supervision
  for Relation Extraction}. In \bibinfo{booktitle}{\emph{Proceedings of the
  52nd Annual Meeting of the Association for Computational Linguistics (Volume
  2: Short Papers)}}. \bibinfo{publisher}{Association for Computational
  Linguistics}, \bibinfo{pages}{732--738}.
\newblock
\urldef\tempurl%
\url{https://www.aclweb.org/anthology/P14-2119/}
\showURL{%
\tempurl}


\bibitem[\protect\citeauthoryear{perspective}{perspective}{2020}]%
        {60_Perspective}
\bibfield{author}{\bibinfo{person}{Jigsaw perspective}.}
  \bibinfo{year}{2020}\natexlab{}.
\newblock \bibinfo{booktitle}{\emph{Perspective API}}.
\newblock
\urldef\tempurl%
\url{https://www.perspectiveapi.com}
\showURL{%
Retrieved Dec 1, 2020 from \tempurl}


\bibitem[\protect\citeauthoryear{Phan}{Phan}{2020}]%
        {59_Phan}
\bibfield{author}{\bibinfo{person}{Trung~T. Phan}.}
  \bibinfo{year}{2020}\natexlab{}.
\newblock \bibinfo{booktitle}{\emph{For the very first time, Reddit revealed
  its user numbers}}.
\newblock
\urldef\tempurl%
\url{https://thehustle.co/12032020-reddit-user-num/}
\showURL{%
Retrieved Dec 1, 2020 from \tempurl}


\bibitem[\protect\citeauthoryear{Policy}{Policy}{2018}]%
        {41_Twitter}
\bibfield{author}{\bibinfo{person}{Twitter~Public Policy}.}
  \bibinfo{year}{2018}\natexlab{}.
\newblock \bibinfo{booktitle}{\emph{Evolving our Twitter Transparency Report:
  expanded data and insights}}.
\newblock
\urldef\tempurl%
\url{https://blog.twitter.com/official/en_us/topics/company/2018/evolving-our-twitter-transparency-report.html}
\showURL{%
Retrieved Dec 1, 2020 from \tempurl}


\bibitem[\protect\citeauthoryear{Preece and Maloney-Krichmar}{Preece and
  Maloney-Krichmar}{2003}]%
        {32_Preece}
\bibfield{author}{\bibinfo{person}{Jenny Preece} {and} \bibinfo{person}{Diane
  Maloney-Krichmar}.} \bibinfo{year}{2003}\natexlab{}.
\newblock \showarticletitle{Online communities: focusing on sociability and
  usability}.
\newblock \bibinfo{journal}{\emph{Handbook of human-computer interaction}}
  (\bibinfo{year}{2003}).
\newblock


\bibitem[\protect\citeauthoryear{Reddit}{Reddit}{2020}]%
        {47_reddit}
\bibfield{author}{\bibinfo{person}{Reddit}.} \bibinfo{year}{2020}\natexlab{}.
\newblock \bibinfo{booktitle}{\emph{Automoderator}}.
\newblock
\urldef\tempurl%
\url{https://www.reddit.com/wiki/automoderator}
\showURL{%
Retrieved Dec 1, 2020 from \tempurl}


\bibitem[\protect\citeauthoryear{Roberts}{Roberts}{2018}]%
        {29_Roberts}
\bibfield{author}{\bibinfo{person}{Sarah Roberts}.}
  \bibinfo{year}{2018}\natexlab{}.
\newblock \showarticletitle{Digital detritus: 'Error' and the logic of opacity
  in social media content moderation}.
\newblock \bibinfo{journal}{\emph{First Monday}} (\bibinfo{year}{2018}).
\newblock


\bibitem[\protect\citeauthoryear{Roberts}{Roberts}{2016}]%
        {9_Roberts}
\bibfield{author}{\bibinfo{person}{Sarah~T. Roberts}.}
  \bibinfo{year}{2016}\natexlab{}.
\newblock \showarticletitle{Commercial Content Moderation: Digital Laborers'
  Dirty Work}.
\newblock \bibinfo{journal}{\emph{Media Studies Publications}}
  \bibinfo{volume}{12} (\bibinfo{year}{2016}).
\newblock
\urldef\tempurl%
\url{https://ir.lib.uwo.ca/cgi/viewcontent.cgi?article=1012&context=commpub}
\showURL{%
\tempurl}


\bibitem[\protect\citeauthoryear{Rocklage and Fazio}{Rocklage and
  Fazio}{2015}]%
        {65_Rocklage}
\bibfield{author}{\bibinfo{person}{Matthew~D. Rocklage} {and}
  \bibinfo{person}{Russell~H. Fazio}.} \bibinfo{year}{2015}\natexlab{}.
\newblock \showarticletitle{The Evaluative Lexicon: Adjective use as a means of
  assessing and distinguishing attitude valence, extremity, and emotionality}.
\newblock \bibinfo{journal}{\emph{Journal of Experimental Social Psychology}}
  \bibinfo{volume}{56} (\bibinfo{year}{2015}).
\newblock
\urldef\tempurl%
\url{https://doi.org/10.1016/j.jesp.2014.10.005}
\showURL{%
\tempurl}


\bibitem[\protect\citeauthoryear{Rocklage, Rucker, and Nordgren}{Rocklage
  et~al\mbox{.}}{2018}]%
        {66_Rocklage}
\bibfield{author}{\bibinfo{person}{Matthew~D. Rocklage},
  \bibinfo{person}{Derek~D. Rucker}, {and} \bibinfo{person}{Loran~F.
  Nordgren}.} \bibinfo{year}{2018}\natexlab{}.
\newblock \showarticletitle{The Evaluative Lexicon 2.0: The measurement of
  emotionality, extremity, and valence in language}.
\newblock \bibinfo{journal}{\emph{Behavior Research Methods}}
  \bibinfo{volume}{50} (\bibinfo{year}{2018}).
\newblock
\urldef\tempurl%
\url{https://doi.org/10.3758/s13428-017-0975-6}
\showURL{%
\tempurl}


\bibitem[\protect\citeauthoryear{Rolf}{Rolf}{2016}]%
        {20_Rolf}
\bibfield{author}{\bibinfo{person}{David Rolf}.}
  \bibinfo{year}{2016}\natexlab{}.
\newblock \bibinfo{booktitle}{\emph{The Fight for Fifteen: The Right Wage for a
  Working America}}.
\newblock \bibinfo{publisher}{The New Press}.
\newblock


\bibitem[\protect\citeauthoryear{Seering}{Seering}{2020}]%
        {seering2020reconsidering}
\bibfield{author}{\bibinfo{person}{Joseph Seering}.}
  \bibinfo{year}{2020}\natexlab{}.
\newblock \showarticletitle{Reconsidering Self-Moderation: the Role of Research
  in Supporting Community-Based Models for Online Content Moderation}.
\newblock \bibinfo{journal}{\emph{Proceedings of the ACM on Human-Computer
  Interaction}} \bibinfo{volume}{4}, \bibinfo{number}{CSCW2}
  (\bibinfo{year}{2020}), \bibinfo{pages}{1--28}.
\newblock


\bibitem[\protect\citeauthoryear{Seering, Wang, Yoon, and Kaufman}{Seering
  et~al\mbox{.}}{2019}]%
        {35_Seering}
\bibfield{author}{\bibinfo{person}{Joseph Seering}, \bibinfo{person}{Tony
  Wang}, \bibinfo{person}{Jina Yoon}, {and} \bibinfo{person}{Geoff Kaufman}.}
  \bibinfo{year}{2019}\natexlab{}.
\newblock \showarticletitle{Moderator Engagement and Community Development in
  the Age of Algorithms}.
\newblock \bibinfo{journal}{\emph{New Media and Society}}
  (\bibinfo{year}{2019}).
\newblock


\bibitem[\protect\citeauthoryear{Shachaf and Hara}{Shachaf and Hara}{2010}]%
        {shachaf2010beyond}
\bibfield{author}{\bibinfo{person}{Pnina Shachaf} {and} \bibinfo{person}{Noriko
  Hara}.} \bibinfo{year}{2010}\natexlab{}.
\newblock \showarticletitle{Beyond vandalism: Wikipedia trolls}.
\newblock \bibinfo{journal}{\emph{Journal of Information Science}}
  \bibinfo{volume}{36}, \bibinfo{number}{3} (\bibinfo{year}{2010}),
  \bibinfo{pages}{357--370}.
\newblock


\bibitem[\protect\citeauthoryear{Sinders}{Sinders}{2017}]%
        {53_Sinders}
\bibfield{author}{\bibinfo{person}{C Sinders}.}
  \bibinfo{year}{2017}\natexlab{}.
\newblock \bibinfo{booktitle}{\emph{Toxicity and tone are not the same thing:
  Analyzing the new Google API on toxicity, PerspectiveAPI}}.
\newblock
\urldef\tempurl%
\url{https://perma.cc/R9BM-V638}
\showURL{%
Retrieved Dec 1, 2020 from \tempurl}


\bibitem[\protect\citeauthoryear{Sood, Antin, and Churchill}{Sood
  et~al\mbox{.}}{2012a}]%
        {49_Sood}
\bibfield{author}{\bibinfo{person}{Sara Sood}, \bibinfo{person}{Judd Antin},
  {and} \bibinfo{person}{Elizabeth Churchill}.}
  \bibinfo{year}{2012}\natexlab{a}.
\newblock \showarticletitle{Profanity use in online communities}. In
  \bibinfo{booktitle}{\emph{In Proceedings of the SIGCHI Conference on Human
  Factors in Computing Systems}}. \bibinfo{publisher}{ACM}.
\newblock


\bibitem[\protect\citeauthoryear{Sood, Churchill, and Antin}{Sood
  et~al\mbox{.}}{2012b}]%
        {6_Sood}
\bibfield{author}{\bibinfo{person}{Sara~Owsley Sood},
  \bibinfo{person}{Elizabeth~F Churchill}, {and} \bibinfo{person}{Judd Antin}.}
  \bibinfo{year}{2012}\natexlab{b}.
\newblock \showarticletitle{Automatic identification of personal insults on
  social news sites}.
\newblock \bibinfo{journal}{\emph{Journal of the American Society for
  Information Science and Technology 63, 2}} (\bibinfo{year}{2012}),
  \bibinfo{pages}{270--285}.
\newblock
\urldef\tempurl%
\url{https://dl.acm.org/doi/10.1002/asi.21690}
\showURL{%
\tempurl}


\bibitem[\protect\citeauthoryear{Suler}{Suler}{2004}]%
        {suler2004online}
\bibfield{author}{\bibinfo{person}{John Suler}.}
  \bibinfo{year}{2004}\natexlab{}.
\newblock \showarticletitle{The online disinhibition effect}.
\newblock \bibinfo{journal}{\emph{Cyberpsychology \& behavior}}
  \bibinfo{volume}{7}, \bibinfo{number}{3} (\bibinfo{year}{2004}),
  \bibinfo{pages}{321--326}.
\newblock


\bibitem[\protect\citeauthoryear{Team}{Team}{2015}]%
        {42_HN}
\bibfield{author}{\bibinfo{person}{HN~Moderation Team}.}
  \bibinfo{year}{2015}\natexlab{}.
\newblock \bibinfo{booktitle}{}.
\newblock
\urldef\tempurl%
\url{https://news.ycombinator.com/threads?id=dang}
\showURL{%
Retrieved Dec 1, 2020 from \tempurl}


\bibitem[\protect\citeauthoryear{Team}{Team}{2019}]%
        {77_YoutubeTeam}
\bibfield{author}{\bibinfo{person}{The~YouTube Team}.}
  \bibinfo{year}{2019}\natexlab{}.
\newblock \bibinfo{booktitle}{\emph{Our ongoing work to tackle hate}}.
\newblock
\urldef\tempurl%
\url{https://blog.youtube/news-and-events/our-ongoing-work-to-tackle-hate}
\showURL{%
Retrieved Dec 1, 2020 from \tempurl}


\bibitem[\protect\citeauthoryear{TensorFlow}{TensorFlow}{2020}]%
        {26_TensorFlow}
\bibfield{author}{\bibinfo{person}{TensorFlow}.}
  \bibinfo{year}{2020}\natexlab{}.
\newblock \bibinfo{booktitle}{\emph{Text Classification}}.
\newblock
\urldef\tempurl%
\url{https://www.tensorflow.org/tutorials/keras/text_classification}
\showURL{%
Retrieved Dec 1, 2020 from \tempurl}


\bibitem[\protect\citeauthoryear{Twitch}{Twitch}{2021}]%
        {twitch_transparency_report}
\bibfield{author}{\bibinfo{person}{Twitch}.} \bibinfo{year}{2021}\natexlab{}.
\newblock \bibinfo{title}{Transparency Report}.
\newblock
\newblock
\urldef\tempurl%
\url{https://safety.twitch.tv/s/article/Transparency-Reports?language=en_US}
\showURL{%
\tempurl}


\bibitem[\protect\citeauthoryear{Varian}{Varian}{2005}]%
        {61_Varian}
\bibfield{author}{\bibinfo{person}{Hal~R. Varian}.}
  \bibinfo{year}{2005}\natexlab{}.
\newblock \showarticletitle{Bootstrap Tutorial}.
\newblock \bibinfo{journal}{\emph{Mathematica Journal}} (\bibinfo{year}{2005}).
\newblock


\bibitem[\protect\citeauthoryear{Varjas, Talley, Meyers, Parris, and
  Cutts}{Varjas et~al\mbox{.}}{2010}]%
        {varjas2010high}
\bibfield{author}{\bibinfo{person}{Kris Varjas}, \bibinfo{person}{Jasmaine
  Talley}, \bibinfo{person}{Joel Meyers}, \bibinfo{person}{Leandra Parris},
  {and} \bibinfo{person}{Hayley Cutts}.} \bibinfo{year}{2010}\natexlab{}.
\newblock \showarticletitle{High school students’ perceptions of motivations
  for cyberbullying: An exploratory study}.
\newblock \bibinfo{journal}{\emph{Western Journal of Emergency Medicine}}
  \bibinfo{volume}{11}, \bibinfo{number}{3} (\bibinfo{year}{2010}),
  \bibinfo{pages}{269}.
\newblock


\bibitem[\protect\citeauthoryear{Vincent}{Vincent}{2020}]%
        {78_Vincent}
\bibfield{author}{\bibinfo{person}{James Vincent}.}
  \bibinfo{year}{2020}\natexlab{}.
\newblock \bibinfo{booktitle}{\emph{Reddit reports 18 percent reduction in
  hateful content after banning nearly 7,000 subreddits}}.
\newblock
\urldef\tempurl%
\url{https://www.theverge.com/2020/8/20/21376957/reddit-hate-speech-content-policies-subreddit-bans-reduction}
\showURL{%
Retrieved Dec 1, 2020 from \tempurl}


\bibitem[\protect\citeauthoryear{Vitak, Chadha, Steiner, and Ashktorab}{Vitak
  et~al\mbox{.}}{2017}]%
        {vitak_harassment}
\bibfield{author}{\bibinfo{person}{Jessican Vitak}, \bibinfo{person}{Kalyani
  Chadha}, \bibinfo{person}{Linda Steiner}, {and} \bibinfo{person}{Zahra
  Ashktorab}.} \bibinfo{year}{2017}\natexlab{}.
\newblock \showarticletitle{Identifying Women's Experiences With and Strategies
  for Mitigating Negative Effects of Online Harassment}.
\newblock \bibinfo{journal}{\emph{Proceedings of the 2017 ACM Conference on
  Computer Supported Cooperative Work and Social Computing}}
  \bibinfo{number}{CSCW} (\bibinfo{year}{2017}), \bibinfo{pages}{1231–1245}.
\newblock


\bibitem[\protect\citeauthoryear{Vogels}{Vogels}{2021}]%
        {pew_harassment_results}
\bibfield{author}{\bibinfo{person}{Emily Vogels}.}
  \bibinfo{year}{2021}\natexlab{}.
\newblock \bibinfo{title}{The State of Online Harassment}.
\newblock
\newblock
\urldef\tempurl%
\url{https://www.pewresearch.org/internet/2021/01/13/the-state-of-online-harassment/}
\showURL{%
\tempurl}


\bibitem[\protect\citeauthoryear{Weisstein}{Weisstein}{2020}]%
        {62_Weisstein}
\bibfield{author}{\bibinfo{person}{Eric Weisstein}.}
  \bibinfo{year}{2020}\natexlab{}.
\newblock \bibinfo{booktitle}{\emph{Bootstrap Methods}}.
\newblock
\urldef\tempurl%
\url{http://mathworld.wolfram.com/BootstrapMethods.html}
\showURL{%
Retrieved Dec 1, 2020 from \tempurl}


\bibitem[\protect\citeauthoryear{Wiener}{Wiener}{1998}]%
        {85_Wiener}
\bibfield{author}{\bibinfo{person}{David Wiener}.}
  \bibinfo{year}{1998}\natexlab{}.
\newblock \showarticletitle{Negligent publication of statements posted on
  electronic bulletin boards: Is there any liability left after Zeran?}
\newblock \bibinfo{journal}{\emph{Santa Clara L Rev}} (\bibinfo{year}{1998}).
\newblock


\bibitem[\protect\citeauthoryear{Wikipedia}{Wikipedia}{2020a}]%
        {64-kinkaid}
\bibfield{author}{\bibinfo{person}{Wikipedia}.}
  \bibinfo{year}{2020}\natexlab{a}.
\newblock \bibinfo{booktitle}{\emph{Flesch–Kincaid readability tests}}.
\newblock
\urldef\tempurl%
\url{https://en.wikipedia.org/wiki/Flesch\%E2\%80\%93Kincaid_readability_tests}
\showURL{%
Retrieved Dec 1, 2020 from \tempurl}


\bibitem[\protect\citeauthoryear{Wikipedia}{Wikipedia}{2020b}]%
        {63_poisson}
\bibfield{author}{\bibinfo{person}{Wikipedia}.}
  \bibinfo{year}{2020}\natexlab{b}.
\newblock \bibinfo{booktitle}{\emph{Poisson Regression}}.
\newblock
\urldef\tempurl%
\url{https://en.wikipedia.org/wiki/Poisson_regression#:~:text=Poisson\%20regression\%20assumes\%20the\%20response,used\%20to\%20model\%20contingency\%20tables.}
\showURL{%
Retrieved Dec 1, 2020 from \tempurl}


\bibitem[\protect\citeauthoryear{Williams and Cothrel}{Williams and
  Cothrel}{2000}]%
        {33_Williams}
\bibfield{author}{\bibinfo{person}{Ruth~L Williams} {and}
  \bibinfo{person}{Joseph Cothrel}.} \bibinfo{year}{2000}\natexlab{}.
\newblock \showarticletitle{Four smart ways to run online communities}.
\newblock \bibinfo{journal}{\emph{MIT Sloan Management Review}}
  (\bibinfo{year}{2000}).
\newblock


\bibitem[\protect\citeauthoryear{Wohn}{Wohn}{2019}]%
        {Wohn2019}
\bibfield{author}{\bibinfo{person}{Donghee~Yvette Wohn}.}
  \bibinfo{year}{2019}\natexlab{}.
\newblock \showarticletitle{Volunteer Moderators in Twitch Micro Communities:
  How They Get Involved, the Roles They Play, and the Emotional Labor They
  Experience}.
\newblock \bibinfo{journal}{\emph{Proceedings of the 2019 CHI Conference on
  Human Factors in Computing Systems}} (\bibinfo{year}{2019}).
\newblock


\bibitem[\protect\citeauthoryear{Xu, Burchfiel, Zhu, and Bellmore}{Xu
  et~al\mbox{.}}{2013a}]%
        {15_Xu}
\bibfield{author}{\bibinfo{person}{Jun-Ming Xu}, \bibinfo{person}{Benjamin
  Burchfiel}, \bibinfo{person}{Xiaojin Zhu}, {and} \bibinfo{person}{Amy
  Bellmore}.} \bibinfo{year}{2013}\natexlab{a}.
\newblock \showarticletitle{An Examination of Regret in Bullying Tweets}. In
  \bibinfo{booktitle}{\emph{Proceedings of the 2013 Conference of the North
  American Chapter of the Association for Computational Linguistics: Human
  Language Technologies}}. \bibinfo{publisher}{Association for Computational
  Linguistics}, \bibinfo{pages}{697--702}.
\newblock
\urldef\tempurl%
\url{https://www.aclweb.org/anthology/N13-1082/}
\showURL{%
\tempurl}


\bibitem[\protect\citeauthoryear{Xu, Burchfiel, Zhu, and Bellmore}{Xu
  et~al\mbox{.}}{2013b}]%
        {44_Xu}
\bibfield{author}{\bibinfo{person}{Jun-Ming Xu}, \bibinfo{person}{Benjamin
  Burchfiel}, \bibinfo{person}{Xiaojin Zhu}, {and} \bibinfo{person}{Amy
  Bellmore}.} \bibinfo{year}{2013}\natexlab{b}.
\newblock \showarticletitle{An Examination of Regret in Bullying Tweets}. In
  \bibinfo{booktitle}{\emph{In Proceedings of the North American Chapter of the
  Association for Computational Linguistics (NAACL-HLT)}}.
\newblock


\bibitem[\protect\citeauthoryear{Yavuz, Levent, and Bahadir}{Yavuz
  et~al\mbox{.}}{2010}]%
        {84_Yavuz}
\bibfield{author}{\bibinfo{person}{Akbulut Yavuz}, \bibinfo{person}{Sahin~Yusuf
  Levent}, {and} \bibinfo{person}{Eristi Bahadir}.}
  \bibinfo{year}{2010}\natexlab{}.
\newblock \showarticletitle{Contexts of uninhibited online behavior: Flaming in
  social newsgroups on Usenet}.
\newblock \bibinfo{journal}{\emph{Educ Technol Soc}} (\bibinfo{year}{2010}).
\newblock


\bibitem[\protect\citeauthoryear{Zehlike, Bonchi, Castillo, Hajian, Megahed,
  and Baeza-Yates}{Zehlike et~al\mbox{.}}{2017}]%
        {56_Zehlike}
\bibfield{author}{\bibinfo{person}{Meike Zehlike}, \bibinfo{person}{Francesco
  Bonchi}, \bibinfo{person}{Carlos Castillo}, \bibinfo{person}{Sara Hajian},
  \bibinfo{person}{Mohamed Megahed}, {and} \bibinfo{person}{Ricardo
  Baeza-Yates}.} \bibinfo{year}{2017}\natexlab{}.
\newblock \showarticletitle{Fa* ir: A fair top-k ranking algorithm}. In
  \bibinfo{booktitle}{\emph{In: Proceedings of the 2017 ACM conference on
  information and knowledge management}}. \bibinfo{publisher}{ACM}.
\newblock


\bibitem[\protect\citeauthoryear{Zhang, Hugh, and Bernstein}{Zhang
  et~al\mbox{.}}{2020}]%
        {zhang2020policykit}
\bibfield{author}{\bibinfo{person}{Amy~X Zhang}, \bibinfo{person}{Grant Hugh},
  {and} \bibinfo{person}{Michael~S Bernstein}.}
  \bibinfo{year}{2020}\natexlab{}.
\newblock \showarticletitle{PolicyKit: Building Governance in Online
  Communities}. In \bibinfo{booktitle}{\emph{Proceedings of the 33rd Annual ACM
  Symposium on User Interface Software and Technology}}.
  \bibinfo{pages}{365--378}.
\newblock


\bibitem[\protect\citeauthoryear{Zhang, Niu, Ré, and Shavlik}{Zhang
  et~al\mbox{.}}{2012}]%
        {24_Zhang}
\bibfield{author}{\bibinfo{person}{Ce Zhang}, \bibinfo{person}{Feng Niu},
  \bibinfo{person}{Christopher Ré}, {and} \bibinfo{person}{Jude Shavlik}.}
  \bibinfo{year}{2012}\natexlab{}.
\newblock \showarticletitle{Big Data versus the Crowd: Looking for
  Relationships in All the Right Places}. In
  \bibinfo{booktitle}{\emph{Proceedings of the 50th Annual Meeting of the
  Association for Computational Linguistics (Volume 1: Long Papers)}}.
  \bibinfo{publisher}{Association for Computational Linguistics},
  \bibinfo{pages}{825--834}.
\newblock
\urldef\tempurl%
\url{https://www.aclweb.org/anthology/P12-1087/}
\showURL{%
\tempurl}


\bibitem[\protect\citeauthoryear{Zhang, Chang, Danescu-Niculescu-Mizil, Dixon,
  Hua, Thain, and Taraborelli}{Zhang et~al\mbox{.}}{2018}]%
        {zhang2018conversations}
\bibfield{author}{\bibinfo{person}{Justine Zhang}, \bibinfo{person}{Jonathan~P
  Chang}, \bibinfo{person}{Cristian Danescu-Niculescu-Mizil},
  \bibinfo{person}{Lucas Dixon}, \bibinfo{person}{Yiqing Hua},
  \bibinfo{person}{Nithum Thain}, {and} \bibinfo{person}{Dario Taraborelli}.}
  \bibinfo{year}{2018}\natexlab{}.
\newblock \showarticletitle{Conversations gone awry: Detecting early signs of
  conversational failure}.
\newblock \bibinfo{journal}{\emph{arXiv preprint arXiv:1805.05345}}
  (\bibinfo{year}{2018}).
\newblock


\end{thebibliography}
